\documentclass[sigconf]{acmart}
\AtBeginDocument{%
  \providecommand\BibTeX{{%
    \normalfont B\kern-0.5em{\scshape i\kern-0.25em b}\kern-0.8em\TeX}}}
\setcopyright{acmcopyright}
\copyrightyear{2023}
\acmYear{2023}
\acmDOI{XXXXXXX.XXXXXXX}
\acmConference[CCS'23]{the 2023 ACM SIGSAC Conference on Computer and Communications Security}{November 26--30, 2023}{Copenhagen, Denmark}
\acmPrice{15.00}
\acmISBN{978-1-4503-XXXX-X/18/06}

\usepackage{tikz}
\usepackage{amsmath}

\usepackage[misc]{ifsym}
\usepackage{booktabs}
\usepackage{algorithm}
\usepackage{algorithmic}
\usepackage{multirow}
\usepackage{makecell}
\usepackage{caption}
\usepackage{pifont}
\usepackage{subcaption,xspace}
\usepackage{enumitem}
\newcommand\modify[1]{{\color{black} #1}}
\newcommand{\parab}[1]{\vspace{0.01in}\noindent{\bf #1} }
\usepackage{comment}

\newcommand\ye[1]{{\color{black} #1}}
\newcommand\fixme[1]{{\color{black} #1}}
\newcommand\revision[1]{{\color{black} #1}}

\newcommand\modifyy[1]{{\color{black} #1}}

\newcommand{\ours}{RETSINA\xspace}

\begin{document}
\date{}

\title[]{Learning from Limited Heterogeneous Training Data: Meta-Learning for Unsupervised Zero-Day Web Attack Detection across Web Domains}

\author{Peiyang Li}
\authornote{Both authors contributed equally to this research.}
\affiliation{%
  \institution{Tsinghua University \& BNRist}
  \city{}
  \country{}}
\email{li-py23@mails.tsinghua.edu.cn}

\author{Ye Wang}
\authornotemark[1]
\affiliation{%
  \institution{Tsinghua University \& BNRist}
  \city{}
  \country{}}
\email{wangye22@mails.tsinghua.edu.cn}

\author{Qi Li$^{\textrm{\Letter}}$}
\affiliation{%
  \institution{Tsinghua University}
  \city{}
  \country{}}
\email{qli01@tsinghua.edu.cn}

\author{Zhuotao Liu}
\affiliation{%
  \institution{Tsinghua University}
  \city{}
  \country{}}
\email{zhuotaoliu@tsinghua.edu.cn}

\author{Ke Xu}
\affiliation{%
  \institution{Tsinghua University}
  \city{}
  \country{}}
\email{xuke@tsinghua.edu.cn}

\author{Ju Ren}
\affiliation{%
  \institution{Tsinghua University}
  \city{}
  \country{}}
\email{renju@tsinghua.edu.cn}

\author{Zhiying Liu}
\affiliation{%
  \institution{Tencent}
  \city{}
  \country{}}
\email{louiszyliu@tencent.com}

\author{Ruilin Lin}
\affiliation{%
  \institution{Tencent}
  \city{}
  \country{}}
\email{ruilin@tencent.com}

\renewcommand{\shortauthors}{Peiyang Li, et al.}

\begin{abstract}
Recently unsupervised machine learning based systems have been developed to detect zero-day Web attacks, which can effectively enhance existing Web Application Firewalls (WAFs). However, prior arts only consider detecting attacks on specific domains by training particular detection models for the domains. These systems require a large amount of training data, which causes a long period of time for model training and deployment. 
In this paper, we propose \ours, a novel meta-learning based framework that enables zero-day Web attack detection across different \modify{domains} in an organization with limited training data. Specifically, it utilizes meta-learning \modify{to share knowledge across these domains}, e.g., the relationship between HTTP requests in heterogeneous domains, to efficiently train detection models.  
Moreover, we develop an adaptive preprocessing module to facilitate semantic analysis of Web requests across different domains and design a multi-domain representation method to capture semantic correlations between different domains for cross-domain model training. 
We conduct experiments using four real-world datasets on different domains with a total of 293M Web requests. The experimental results demonstrate that \ours outperforms the existing unsupervised Web attack detection methods with limited training data, e.g., \ours needs only 5-minute training data to achieve comparable detection performance to the existing methods that train separate models for different domains using 1-day training data.  
We also conduct real-world deployment in an Internet company. \ours captures on average 126 and 218 zero-day \revision{attack requests} per day in two domains, respectively, in one month.
\end{abstract}




\maketitle


\section{Introduction}

Web services suffer from various Web attacks (e.g., SQL injection). Web Application Firewalls (WAFs) \cite{prandl2015study,ModSecurity,WebKnight,Naxsi} have become the de facto Web attack defense mechanisms against the attacks.  
However, since WAFs detect Web attacks according to manually configured rules, the zero-day Web attacks generating unknown attack patterns can easily evade WAFs. 
Recently 
unsupervised machine learning systems \cite{12vartouni2018anomaly,13park2018anomaly,16tang2020zerowall} have been developed to 
\revision{complement existing WAFs to detect zero-day attacks}. 
These systems learn patterns of benign Web requests and then identify zero-day attacks \revision{whose patterns deviate from benign requests.}

However, these ML-based systems fail to align with the requirements of \modify{Web attack detection systems} in large-scale, operational environments. First, an organization hosting Web services often maintains multiple Web domains for different Web applications. These domains normally exhibit heterogeneity because they enable different Web functionalities\revision{, e.g., Google Scholar and Gmail are two heterogeneous domains hosting different Google applications.} However, existing methods \cite{12vartouni2018anomaly,13park2018anomaly,16tang2020zerowall,6wang2017detecting,8liang2017anomaly,7liu2022deep} only focus on developing the detection model for one specific domain, which are developed independently for one domain and can hardly be generalized to detect attacks on other domains. 
Second, the detection model requires frequent model retraining to avoid significant performance degradation \cite{cretu2008casting,cretu2009adaptive,han2021log,pendlebury2019tesseract,jan2020throwing} due to \modify{the} Web services update \cite{zheng2012investigating}. \revision{\revision{According to our study on real-world Web services, we find that most of the services are regularly updated, e.g., the average update interval of four real-world services in our study is only 3.79 days.}}
It is difficult to retrain and update detection models due to \revision{the} required long period of time of training data collection \cite{16tang2020zerowall}.
In addition, collecting such a large amount of production data incurs significant data preprocessing or privacy \revision{breach} \cite{liu2021privilege}.

In this paper, we develop RETSINA, a novel meta-learning based framework for zero-day Web attack detection across multiple domains. 
\ours utilizes meta-learning to jointly \ye{train detection models for new deployed domains based on limited heterogeneous data.}   
\ours exploits correlations between heterogeneous requests generated from various domains to build a universal detection model. 
Based on the universal detection model and limited requests from each domain,  
\ours generates a separate domain-specific detection model to detect attacks for the domain, which realize efficient model training and update for the domain.

However, it is challenging to utilize meta-learning to achieve zero-day Web attack detection for multiple domains. 
First, it is difficult to extract valid information from such requests due to the heterogeneity and complexity of processing HTTP requests across heterogeneous domains. 
Second, 
since lexical features of HTTP requests are not in the same semantic space among different domains \cite{6wang2017detecting,12vartouni2018anomaly,16tang2020zerowall,zhang2017deep}, we cannot easily capture feature correlations across Web domains.
Third, it is not easy to transfer \ye{its} knowledge to the domain-specific detection models according to the universal detection model 
\modify{because the structures of detection models vary for different domains.} 

In order to address these issues, we develop three key designs to enable 
knowledge sharing across heterogeneous domains. First, we develop an adaptive approach to facilitate semantic analysis of HTTP requests. It can automatically generate domain-specific strategies to eliminate redundant information that is not useful for attack detection and then produce valid tokens for training machine learning models. 
Second, we propose a feature representation approach for multi-domain requests. It performs token alignment operations by utilizing orthogonal transformations to measure the similarity of tokens across domains. Based on this similarity, it maps all tokens to the same feature space, which facilitates machine learning models to capture correlations and share knowledge across heterogeneous domains. 
\modify{Third, we develop a dedicated two-loop strategy for training the universal detection model, which updates parameters that can be inherited by the target domain.  
This ensures that the knowledge learned by the universal detection model is fully transferable to different target domains.}

We evaluate \ours using four real-world datasets collected from four different domains that provide different types of Web services to demonstrate the effectiveness of \ours. \modify{In particular, we conduct a comparative analysis of RETSINA against two state-of-the-art unsupervised Web attack detection methods \revision{\cite{16tang2020zerowall, xu2021deep}}, one of which is a knowledge-sharing-based framework that is proposed for multi-task security problems, and two supervised Web attack detection techniques \cite{zhang2017deep,zhou2016attention}.}
The experimental results demonstrate that RETSINA outperforms the existing methods with limited training data, e.g., RETSINA needs only 5-minute training data to achieve comparable detection performance for different domains to the existing methods that train separate models for different domains using 1-day training data. We perform ablation experiments to analyze how different modules of \ours contribute to the detection. We also deploy \ours in real-world production services in 
an Internet company. \ours detects on average 126 and 218 zero-day \revision{attack requests} per day in two domains, respectively.
 
In summary, we make the following contributions:
\begin{itemize}[leftmargin=*, noitemsep, topsep=0pt]

    \item We propose \ours, an unsupervised framework to detect zero-day Web attacks by utilizing meta-learning, which is the first one to detect Web attacks across multiple domains with limited and heterogeneous data.
    \item We design an adaptive method to preprocess multi-domain Web requests, which automatically converts unstructured raw requests into structured token sequences for training and detection. 
    \item We develop a multi-domain representation method to construct unified feature representations across different Web domains, which can be easily extended to enable detection for new domains.
    \item We conduct extensive evaluations 
    with a total of 293M Web requests from 4 real-world domains owned by an Internet company. Evaluation results demonstrate the effectiveness of \ours for multiple heterogeneous Web domains with limited training data.
    \item We perform real-world deployment in real production and evaluate the performance for one month, detecting on average 126 and 218 zero-day \revision{attack requests} per day, demonstrating the superiority of \ours with respect to effectiveness and robustness.
\end{itemize}

\section{Problem Statement}
\label{sec:problem}

We consider the scenario where an Internet company operates multiple frequently updated domains and consistently releases new domains. 
These domains exhibit heterogeneity since they have different functionalities to support different services. 
Meanwhile, these domains follow the same development specification and can share the same underlying interface (e.g., the authentication interface). 
Web attack detection models are needed for updated or newly developed domains. However, 
to keep up with the rapid domain release, only a limited number of data is available for these domains.

Our goal is to develop zero-day Web attack detection models for each \textit{target domain} 
with limited training data, by leveraging the heterogeneous data collected from several \textit{auxiliary domains}. 
While requests from the target domain are limited, each auxiliary domain has adequate request data.
\revision{The developed detection model will work together with existing rule-based WAFs, i.e., detecting zero-day attacks that evade WAFs.}

\begin{figure*}[t] 
    \centering 
    \includegraphics[width=0.95\linewidth]{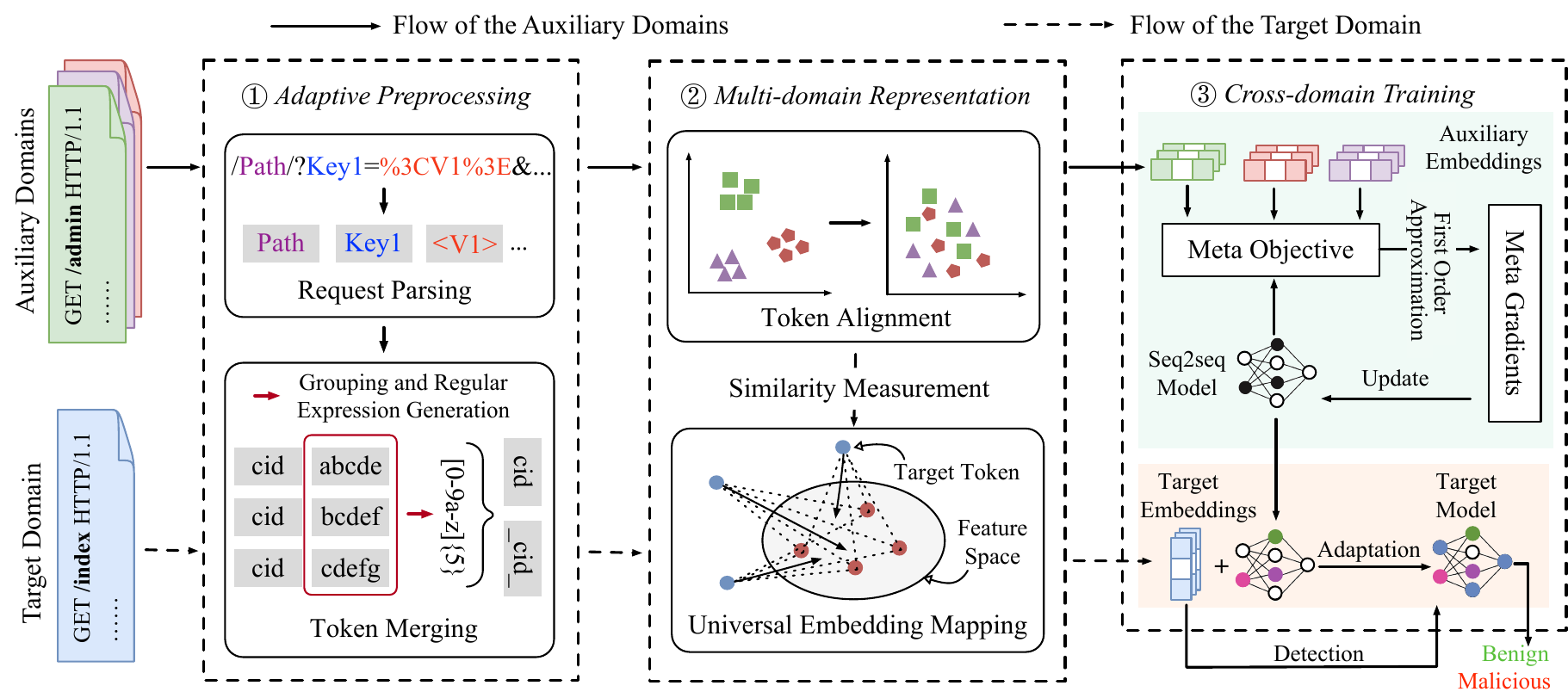}
    \caption{The overview of \ours.} 
\label{fig:framework} 
\end{figure*}

Similar to existing approaches (e.g., \cite{zhang2017deep,16tang2020zerowall}),  
\revision{we focus on analyzing URLs as well as the message bodies with the ``x-www-form-urlencoded'' content type, which covers a majority of Web attacks. To validate this issue, we manually analyze 
352 Web vulnerabilities collected from a popular GitHub project \cite{xray}, and find that 83.2\% of these vulnerabilities can be detected by \ours.}
We do not consider the other types of message bodies since their formats are not restricted by the HTTP protocol. 
We also skip headers 
because they have limited information for attack detection, and more crucially, anomalies in headers can be identified by certain rules defined in traditional WAFs.

In this paper, we leverage the state-of-the-art unsupervised method, i.e., ZeroWall \cite{16tang2020zerowall}, \fixme{as the basic detection model in our framework}. \revision{Zerowall leverages an embedding layer and a sequence-to-sequence (seq2seq) model to learn the benign pattern of Web requests by minimizing the reconstruction error between the input benign requests and the output reconstructed requests. Then it detects malicious Web requests whose patterns deviate from the benign pattern.}
\revision{In particular, the embedding layer converts each input sequence of requests into a vector sequence. The seq2seq model consists of a sequence-based encoder-decoder network that encodes a vector sequence into the latent feature space and decodes it back to a new vector sequence, and a generator that converts the vector sequence into the discrete output sequence. }

\section{Framework Overview}
\label{sec:framework}

We propose RETSINA for zero-day Web attack detection across heterogeneous domains with limited data. Our key insight is that semantic correlations exist between requests from different heterogeneous domains. 
\revision{The rationale behind semantic correlations is that domains operated by the same company follow the same development specification and share the same underlying interface. For example, the corporate data we collected for experiments established a series of cross-business parameter specifications for data governance. This is prevalent in other organizations. For example, multiple Google Web services including Google Cloud use five specific URL parameters with the prefix ``utm’’ \cite{utm} to track user traffic.}
This insight indicates that the detection on one domain can benefit from the existing detection experiences from other domains.

Based on this insight, \ours leverages meta-learning to gain detection experience over models from multiple auxiliary domains. Since meta-learning learn patterns and relationships that are common across different domains, \ours~\revision{is} capable of training the detection models on target domains with limited data. \ours consists of the following three modules.

\begin{itemize}[leftmargin=*,noitemsep,topsep=0pt]
\item \textbf{Adaptive Preprocessing.} Adaptive preprocessing converts each unstructured request into a structured token sequence, to facilitate semantic analysis during model training and detection. 
\modifyy{It} first parses requests into token sequences based on punctuation and then merges 
 \modifyy{tokens with inessential information according to the strategy 
automatically generated} \revision{for} each domain. 

\item \textbf{Multi-domain Representation.} Multi-domain representation projects tokens from heterogeneous domains into the same feature space in which the tokens with similar semantics are close. It chooses a domain as the base domain, and tokens in other domains are represented as a weighted sum of tokens in the base domain according to their semantic similarities. 

\item \textbf{Cross-domain Training.} Cross-domain training obtains a detection model for each target domain with limited training data. We first build a universal initial model using data from auxiliary domains and then adapt the model to the target domain using the limited data from that domain. Particularly, the universal initial model is trained to be adapted well to new domains by leveraging the idea of meta-learning.

\end{itemize}

Figure \ref{fig:framework} shows the workflow of our framework from the perspective of auxiliary domains and target domains. For auxiliary domains, the adaptive preprocessing module converts requests from each domain into token sequences. Then the multi-domain representation module chooses a base domain among all domains and builds embeddings for all tokens according to the base domain. Using these embeddings, the cross-scenario training module obtains a universal initial model. 
For each target domain, the adaptive preprocessing module converts requests from this domain into token sequences. The multi-domain representation module builds embeddings for tokens in this domain according to the chosen base domain. Using these embeddings, in the cross-scenario training module, the universal initial model is then adapted to a target domain-specific model, on which we can perform online zero-day Web attack detection. 
It is worth noticing that, 
\ours finishes the computation of auxiliary domains in advance, without knowing the target domain, thus enabling the detection model development for the target domain with limited data as well as a short development time.

\section{Design Details}
\label{sec:detail}
In this section, we present detailed \revision{designs} of \ours.

\subsection{Adaptive Preprocessing}
\begin{figure}
    \centering
    
    \includegraphics[width=8cm]{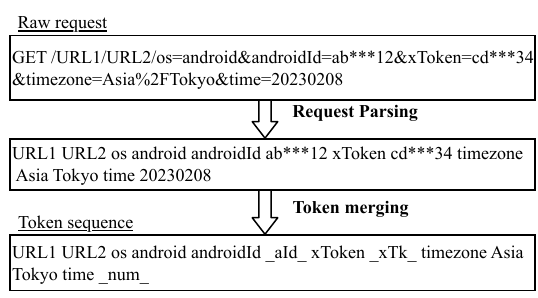}
    \vspace{-0.3cm}
    \caption{One example of adaptive preprocessing.}
    \label{fig:pre_example}
\end{figure}

Adaptive preprocessing aims at converting HTTP requests into token sequences, to facilitate semantic analysis during model training and attack detection. As mentioned earlier, HTTP requests are strings with complex semantics that \modify{vary} across different domains. Therefore, adaptive preprocessing automatically generates domain-specific preprocessing strategies, to obtain token sequences with accurate semantics.
In particular, for each domain, each request is first parsed into a token sequence. 
Then, by extracting the format of tokens, a preprocessing strategy, named as merging strategy, is automatically generated. It is applied to merge a number of tokens \modifyy{into fixed tokens.} An example is shown in Figure \ref{fig:pre_example}.

\parab{Request Parsing.} We parse each request into several parts according to the HTTP protocol \revision{and convert letters to lowercase.} Specifically, \modifyy{we decode and parse a request} into a path and multiple pairs of a key and a value in a query. 
Then we decode and split each part into tokens according to the punctuation (e.g., `/’) and space.

\parab{Token Merging.} 
We now merge tokens with similar semantics into the same tokens to obtain tokens with accurate semantics. We observe that frequently, the values of a key can be uncertain, and may even evolve with time, e.g., the number of \revision{distinct} values of key ``$\texttt{androidId}$” can be extremely huge. 
These values complicate the HTTP requests, but provide no additional benefits to attack detection since they have almost the same semantics. We regard these values as \textit{inessential} tokens and merge them as the same tokens.
In particular, we perform the following steps to merge these inessential tokens: i) We first group values by their keys to obtain each key's values. We identify keys whose number of distinct values is large, then for each such key, a regular expression is generated to match all its values for merging. Notably, the regular expression should be as small as possible 
\revision{to extract accurate semantics best.}
For example, for a set of values that only contains numbers, ``\texttt{[0-9]+}'' is generated instead of ``\texttt{[0-9a-z]+}''.
For a simple implementation, we devise several common-used regular expressions, such as for numbers, and conduct auto selection among them. 
ii) After obtaining the regular expression of a key, for each of its values, we merge the value token in the token sequence using the same token (e.g., ``$\texttt{\_aId\_}$'' for key ``$\texttt{androidId}$'' in Figure \ref{fig:pre_example}) if the value matches its regular expression. Please refer to Appendix \ref{sec:cs_ap} for more details about this process. \looseness=-1%

In addition, we \revision{replace} all low-frequency tokens using the same token ``$\texttt{\_other\_}$'', to allow our preprocessing to deal with previously unseen tokens.
\revision{Note that the ``$\texttt{\_other\_}$'' token category generally implies that the requests are anomalous, since the low-frequency tokens with benign values have been merged according to our regular expressions.
} 
Given a collection of HTTP requests as our training data, besides identifying the regular expression during token merging, we also identify a \textit{token set} which contains all tokens in the token sequences after request parsing and token merging. When new HTTP requests come, we parse these requests and merge tokens according to the regular expression. We also \revision{replace} tokens that are not in the token set using ``$\texttt{\_other\_}$''.

Our preprocessing automatically generates preprocessing strategies for each domain to merge inessential tokens. 
While accurately extracting semantics for each domain, it can be easily adapted to new domains without expertise effort.
However, existing approaches use general preprocessing strategies for all domains, extracting inadequate semantics to differentiate abnormal when the strategies are too loose \cite{8liang2017anomaly,16tang2020zerowall}, or retaining overly complex semantics to be learned and generalized when the strategies are too strict~\cite{zhang2017deep,13park2018anomaly,12vartouni2018anomaly}.  We provide a detailed example for comparison in Appendix \ref{sec:cs_ap}.

\subsection{Multi-domain Representation}
\label{sec:multi_repre}
We now transform preprocessed token sequences into representations suitable for detection. 
Word embedding techniques use real-valued vectors to represent each token in such a way that tokens with similar semantics are expected to be closer in the continuous vector space \cite{mikolov2013efficient}. Compared to conventional representations like one-hot embedding, they produce lower-dimensional embeddings with semantic similarity and are thus widely applied in natural language processing. However, \revision{tokens from different domains will be separately projected into different vector spaces\revision{\footnote{\revision{Note that combining all tokens from multiple domains in one set is capable of producing representations in the same vector spaces. However, such a method cannot be extended to new domains because it requires obtaining representations of all domains at the same time. 
}}},} which overlooks the semantic similarity between different domains, rendering the knowledge distilling from one domain incapable to improve detection on other domains. 

With the expectation that similar tokens across all domains can have close representations, we propose multi-domain representation, 
where tokens from any domain will be represented as a weighted sum of a universal token set.
We observe that, if token sets of two domains overlap on a token, this overlapping token \revision{generally} has similar semantics in both domains. \revision{
The following three cases can lead to semantics similarity in overlapping tokens. i) Domain developers follow the same development specifications, e.g., two domains share several overlapping tokens when using the same API. ii) Developers use common tokens to express certain concepts. For example, even without development specifications, ``\texttt{version}'' typically describes a set of numbers identifying an update. Such tokens rarely incur semantic overloading. iii) Our adaptive preprocessing merges similar values into the same predefined tokens that represent the same semantics across different domains. Our empirical study on real-world data also confirms this observation. For example, we manually investigate the overlapping tokens between the two domains in our experiment data and find that 120 of the 145 overlapping tokens share similar semantics in both domains. }

Based on the observation above, intuitively, we select one base domain and then, 
all these overlapping tokens shared between the base domain and each other domain can serve as the references for aligning the tokens of them. \revision{Note that we leverage orthogonal transformations to align tokens (see Equation (\ref{eq.linear_transformation})) so that a few overlapping tokens that do not share the same semantics have negligible impacts on token alignment \cite{smith2017offline}. }
Then, the similarity between any pair of tokens can then be computed and used to represent each token using the weighted sum of the universal token set (i.e., the token set of the base domain). Below we describe the details. Important notations are summarized in Table \ref{tab:notion}.  

\begin{table}[t]
\centering
\caption{Notations}
\begin{tabular}{@{}c|l@{}}
\toprule
\textbf{Notation} & \textbf{Explanation} \\ \midrule
\revision{$Q^m$} & The token set of the $m_{\rm th}$ domain\\
$T^m$ & The token sequences of the $m_{\rm th}$ domain\\
${T}^m_i$ & The $i_{th}$ token sequence of ${T}^m$ \\ 
$D^m$ & HTTP requests of the $m_{\rm th}$ domain \\\bottomrule
\end{tabular}
\label{tab:notion}
\end{table}

\parab{Token Alignment.} We first align the tokens in different domains according to their semantic similarities. 
Among all $M$ domains, we select a domain $u$ as our base domain and align the tokens in each other domain with tokens in the base domain. 
Given the token sequences, 
we first compute the preliminary representations $\{W^i\}^M_{i=1}$ for \revision{tokens $Q^i$ in}  each domain \revision{$i$} separately. Here we adopt Word2vec \cite{mikolov2013efficient} which mines contextual information within token sequences by predicting missing tokens from a surrounding window.
Then for each domain $v$, we find a \revision{orthogonal} linear transformation to project the preliminary representation of $v$ to $W^u$ vector space, 
\begin{equation}
    W^{v \rightarrow u} = O_v \cdot W^v,  ~ s.t. O_v^TO_v = I,
    \label{eq.linear_transformation}
\end{equation}
where $I$ denotes the identity matrix. In other words, $O_v^T$ can project tokens in domain $u$ back into $W^v$ vector space of domain $v$.
We find the transformation by minimizing the squared reconstruction error on the overlapping tokens \revision{$X_{uv}$} as follows:
\begin{equation}
    \underset{O_v}{min} \sum_{x\in \revision{X_{uv}}}^{p} \left\| W^u(x) - O_v \cdot W^v(x) \right\|^2,  ~ s.t. O_v^TO_v = I,
    \label{eq.objective}
\end{equation}
Note that, we can easily identify these overlapping tokens by finding the intersection of token sets, \revision{i.e., $X_{uv}=Q^u\cap Q^v$}, without resorting to manual effort.
Unfortunately, Equation (\ref{eq.objective}) cannot be optimized by simply applying gradient descent methods because of the orthogonal constraint for $O_v$. Instead, we use the singular value decomposition 
based method \cite{smith2017offline} under the condition that $W^u$ and $W^v$ are normalized. Readers can refer to the original paper for more details.

\parab{Universal Embedding Mapping.} We now map tokens from different domains to a universal embedding space. 
According to the Equation (\ref{eq.linear_transformation}) which aligns the preliminary representations of domain $v$ with those of the base domain $u$, the similarity between a token $t_v$ in domain $v$ and a token $t_u$ in domain $u$ can be computed following the equation below:
\begin{equation}
    S(t_v,t_u)=W^u(t_v) \cdot A \cdot W^{v \rightarrow u}(t_u),
\end{equation}
where A is a learnable parameter initialized as an identity matrix \revision{and is jointly trained with the seq2seq model}.
With the token similarity, each token can be represented as the weighted sum of the token set \revision{$Q^u$} of the base domain \revision{ $u$, named as \textit{universal embedding}}:
\begin{equation} \label{eq:part_ur}
    U^v(t_v) = \sum_{t_i\in Q^u} \frac{e^{S(t_v,\revision{t_i)}}}{\sum_{t_j\in Q^u} e^{S(t_v,\revision{t_j})}} \cdot E(\revision{t_i})
\end{equation}
where the multiplicator on the left is the weight calculated by normalizing the similarity using a softmax function, and $E$ is a matrix whose shape is $\left\| \revision{Q^u}  \right\| \times d$. We initialize $E$ randomly and \revision{train it jointly with the seq2seq model. }

Besides, \revision{a domain-specific embedding term} is integrated to support representing the tokens derived from URL paths, e.g., token "CGI1" from "/URL1/URL2/CGI1\dots". Note that, these tokens are \revision{domain-}specific and have no similar context structure with other tokens, thus can only be ``memorized'' by the models. \revision{Finally, by summing up the universal embedding $U^v$ in Equation (\ref{eq:part_ur}) and the domain-specific embedding $E^v$, our \textit{multi-domain representation} for domain $v$ is computed:}
\begin{equation}
    \tilde{U}^v(x) = U^v(x) + E^v(x),
    \label{eq.final_embedding}
\end{equation}
where $E^v$ is a matrix whose shape is $\left\| \revision{Q^v} \right\| \times d$. 
$E^v$ is initialized as a zero matrix and is trained only to obtain the detection model for a specified target domain.

\begin{figure}[t] 
    \centering 
    \includegraphics[width=0.85\linewidth]{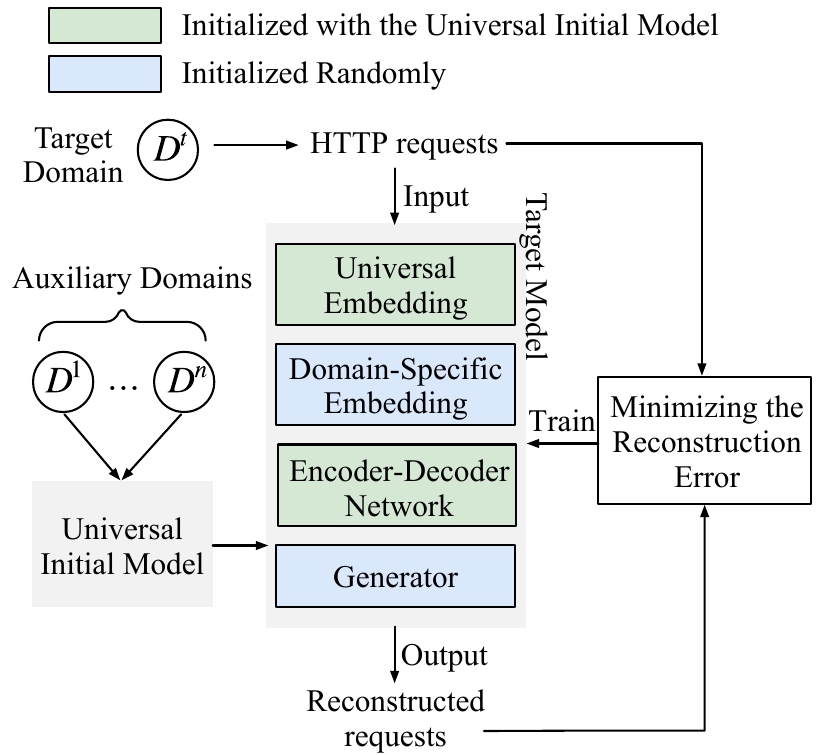}
    \vspace{-0.1cm}
    \caption{Training of the target model.} 
\label{fig:model} 
\end{figure}

\begin{algorithm}[t]
	\caption{Training the universal initial model}
	\begin{algorithmic}[1]
        \small
            \REQUIRE
            \revision{Model $f_{\theta}$}; token sequences of $M$ domains $T_M=\{T^1,T^2,\dots,T^M\}$; learning rate $\alpha$; step size $s$.\\
	    \ENSURE
            \revision{Parameters ${\theta}^*$ of universal} initial model.
	    \STATE{Randomly initialize $\theta$;}
		\WHILE{$f_\theta$ is not convergent}
            \STATE{${\theta_{temp}} \leftarrow \theta$;}
		    \STATE{$T^k \leftarrow$ Sample a domain $k$ from $T_M$;}
               \FOR{$j = 1,2,...,s$}
		            \STATE{$\revision{b^k} \leftarrow$ Sample a batch data from $T^k$;}
		            \STATE{Update $\theta$ according to Eq. (\ref{eq:inner}) with \revision{$b^k$};}
    		    
		    \ENDFOR
        \STATE{$U_{T^k}(\theta) \leftarrow \theta$ }
        \STATE{$\revision{b^k} \leftarrow$ Sample a batch data from $T^k$;}
		\STATE{Compute $g_{meta}$ according to Eq. (\ref{eq:g_outer_appr}) with \revision{$b^k$} and $U_{T^k}(\theta)$;}
        \STATE{$\theta \leftarrow {\theta_{temp}}$;}
		\STATE{update $\theta$ according to Eq. (\ref{eq:outter}) with $g_{meta}$;}
		\ENDWHILE
		\STATE{${\theta}^* \leftarrow \theta$;}
		\RETURN ${\theta}^*$
		
	\end{algorithmic}
	\label{algo:construct2}
\end{algorithm}

\subsection{Cross-domain Training}

Cross-domain training obtains a detection model for each target domain by combining the knowledge from both the target domain and the other auxiliary domains. 
It utilizes meta-learning to better exploit the knowledge from auxiliary domains.
Specifically, we first build a universal initial model using training data from multiple auxiliary domains, with the expectation that it exploits several auxiliary domains, yet provides better performance on any novel domain without resorting to a large amount of training data. To obtain the detection model for a specified target domain, the universal initial model is then adapted using the limited training data from that domain.

\parab{Universal Initial Model.} 
Using the training data from multiple domains, we produce a model
with parameters that can be adapted well to new domains by leveraging the idea of meta-learning \cite{finn2017model}. 
\revision{
Note that, though the meta-learning algorithm \cite{finn2017model} is designed for supervised learning, it can be also applied to our unsupervised task that uses unlabeled data. This is because the objective of our task is to identify attacks by reconstructing each request, which is a supervisory signal. 
}
\revision{Formally, the loss $\mathcal{L}_{T^i}$ on training data $T_i$ from domain $i$ is defined as:
\begin{equation}
    \mathcal{L}_{T^i}(\theta) =\sum_{T^i_j\in T^i} R(T^i_j,f_{\theta}(T^i_j)),
    \label{eq:base_loss}
\end{equation}
where $f_{\theta}$ is our model with the learnable parameters $\theta$, $T^i_j$ is the $j$-th token sequence of the domain $i$, and $R(\cdot)$ is the negative log-likelihood function, i.e., reconstruction error. Note that $f_{\theta}$ is composed of the multi-domain representation and the seq2seq model, and these two parts are optimized jointly. The details of the multi-domain representation are described in Section \ref{sec:multi_repre}, and the seq2seq model, which consists of the encoder-decoder network and the generator, uses the same design as ZeroWall \cite{16tang2020zerowall}.} 

\revision{Given training data $T_M=\{T^1, T^2, \dots, T^M\}$ from M domains, our} objective function, denoted as the meta objective, aims to find the following parameters $\theta^*$:

\begin{equation}
    \theta^* = \underset{\theta}{argmin} \sum_{T^i\in T_M} \mathcal{L}_{T^i}(U_{T^i}(\theta)),
    \label{eq:para_obj}
\end{equation}
where $U_{T^i}(\cdot)$ is the parameter update operator using gradient \revision{descents} on the training data of $i$-th domain. 
Equation (\ref{eq:inner}) shows a step in the gradient \revision{descents}:
\begin{equation}
    \theta ^{'} = \theta - \alpha \nabla_{\theta} {\mathcal{L}_{T^i}(\theta)},
    \label{eq:inner}
\end{equation}
where $\alpha$ is a hyperparameter representing the learning rate. $U_{T^i}(\theta)$ may contain several steps in Equation (\ref{eq:inner}), but below we let $U_{T^i}(\theta)$ contain only one step for the simplicity of discussion, i.e., $U_{T^i}(\theta)=\theta ^{'}$.
Intuitively, the meta objective ensures that a model initialized with $\theta^*$ can be trained (using gradient \revision{descents}) to achieve the minimized loss under each domain.

We \modifyy{leverage} gradient \revision{descents} to solve this meta objective and \modifyy{denote the gradient for optimizing it as} the meta gradient $g_{meta}$:
\begin{equation}
\begin{split}
    g_{meta} = & \sum_{T^i} \nabla_{\theta} \mathcal{L}_{T^i}(U_{T^i}(\theta)) \\
    = & \sum_{T^i} \nabla_{U_{T^i}(\theta)} \mathcal{L}_{T^i}(U_{T^i}(\theta)) \nabla_{\theta} U_{T^i}(\theta) \\
    = & \sum_{T^i} \nabla_{U_{T^i}(\theta)} \mathcal{L}_{T^i}(U_{T^i}(\theta)) \nabla_{\theta} (\theta - \alpha \nabla_{\theta} {\mathcal{L}_{T^i}(\theta)})\\
    = & \sum_{T^i} \nabla_{U_{T^i}(\theta)} \mathcal{L}_{T^i}(U_{T^i}(\theta)) (I - \alpha \nabla_{\theta} \nabla_{\theta} {\mathcal{L}_{T^i}(\theta)})\text{.}
    \label{eq:g_outer}
\end{split}
\end{equation}
Since the high-order derivatives in $g_{meta}$ are expensive to compute, we apply the first-order approximation \cite{nichol2018first} to the meta gradient, yielding the following approximation: 
\begin{equation}
    g_{meta} \approx \sum_{T^i} \nabla_{U_{T^i}(\theta)} \mathcal{L}_{T^i}(U_{T^i}(\theta)).
    \label{eq:g_outer_appr}
\end{equation}

After obtaining $g_{meta}$, a naive way to optimize Equation (\ref{eq:para_obj}) is to perform gradient \revision{descents} iteratively according to $g_{meta}$. However, as shown in Figure \ref{fig:model}, in our model
the domain-specific embedding and the generator that creates the transformation between tokens 
and the vector space needs to be reinitialized for processing new domains. Therefore, the learned knowledge represented in these parameters cannot be retained \cite{you2020co} when training on new domains. 

To address this issue, we do not update the parameters of domain-specific embedding and generator when applying $g_{meta}$, which enables us to ``embed'' the knowledge to the parameters that can be retained. 
Formally, the universal initial model is updated as follows:
\begin{equation}
    \theta \leftarrow \theta - \alpha \cdot g_{meta} \odot \boldsymbol{K},
    \label{eq:outter}
\end{equation}
where $\odot$ represents the element-wise multiplication and $\boldsymbol{K}$ is a parameter mask whose values are 0 for the parameters of the domain-specific embedding and the generator, and 1 otherwise.

Algorithm \ref{algo:construct2} shows the pseudocode of training the universal initial model. 
We first randomly initialize $\theta$ (line 1). Then we apply a two-loop strategy, consisting of an inner and an outer loop, to iteratively update $\theta$. In lines 4-9, the inner loop updates $\theta$ using a randomly selected domain $k$ and obtain $U_{T^i}(\theta)$, which simulates model training on any given target domain. In lines 10-13, the outer loop updates $\theta$ by computing $g_{meta}$. The training data for both the inner loop and outer loop are sampled independently from the randomly chosen domain. It is worth noting that, unlike the outer loop which does not update domain-specific parameters, all parameters are still eligible for updates in the inner loop.

\parab{Target Model.} \revision{We take the universal initial model as the starting point to obtain a detection model for the target domain (denoted as the target model). Note that the structure of the target model is the same as that of the universal initial model. Specifically, the target model inherits the parameters of universal embedding and the encoder-decoder network from the universal initial model, and reinitializes the domain-specific embedding and generator. Given limited data from the target domain, all learnable parameters of the target model are jointly optimized by minimizing the loss function defined in Equation (\ref{eq:base_loss}).} \looseness=-1

\section{Evaluation}
\label{sec:evaluation}

\subsection{Settings} \label{sec:setting}

\parab{Datasets.} As shown in Table \ref{tab:datasets}, we collect four datasets from different domains provided by a world-leading Internet company: 1) OV is an online video platform for streaming media; 2) SNS provides social networking service; 3) SF is a sports forum; 
and 4) IdM offers an identity management service.
All four domains serve different applications, thus providing good diversity for our evaluation across heterogeneous domains. Specifically, we collect the HTTP requests allowed by the company's signature-based WAF for two consecutive days for each domain. The data from the first and second days are used for training and testing, respectively. 
To collect limited training data, we select consecutive 5-minute data starting from the same time from the total (1-day) training dataset of each domain. \revision{For ease of data collection, we do not consider the intervals between the data collection for the auxiliary domains and the target domain. However, such intervals do not impact the model performance, because the semantic correlations between different domains are derived from the same development specifications that do not change over time. We confirm this in real-world deployment experiments (see Section \ref{sec:real-world}).}

\parab{Ground truth.} The security operators collect the ground truth of attacks in the testing set using the following two approaches. 
\revision{First, they manually perform Web log analysis by investigating specific predefined keywords generated according to prior detected events. Second, they \modifyy{examine} all requests detected by \ours and the baselines. A request is identified as an attack if \modifyy{it i) includes malicious payloads, e.g., a code snippet used for injection, or ii) does not follow the normal user behavior of Web applications, e.g., attempting} to access administration interfaces.}
The rest of the requests are considered benign. 
\revision{Notably, the mislabeled attacks, if any, in our ground truth may lead to a higher reported recall, but they will not impact the fairness of comparisons with baselines.}

\parab{Metrics. } We evaluate the detection performance using precision (Pre), recall (Rec), and F1-score (F1), which are also widely used in previous work since zero-day Web attacks only comprise a very small proportion of all HTTP requests.
 
\parab{Baselines.} We choose the two unsupervised anomaly detection methods ZeroWall and MTL. We also select the two supervised techniques SCNN and SRNN, to evaluate whether using some known attack requests yields a better detection system.

\begin{itemize}[leftmargin=*, noitemsep, topsep=0pt]
    \item \textbf{ZeroWall.} ZeroWall is the state-of-the-art unsupervised zero-day Web attack detection method. It parses requests according to punctuation and leverages a seq2seq model to achieve end-to-end detection \cite{16tang2020zerowall}. We also test several widely-used unsupervised zero-day Web attack detection methods. For example, SAE \cite{12vartouni2018anomaly} uses n-gram for tokenization, then leverages the stacked auto-encoder and the isolation forest for feature extraction and detection. However, these methods perform very poorly on our real-world datasets, and the F1-scores are almost 0. Similar results have also been reported in previous work \cite{16tang2020zerowall}. 
    \item \textbf{MTL.} 
    \revision{Neither ZeroWall nor any existing methods for detecting Web attacks can utilize data from auxiliary domains. To enable more fair comparisons, we include a knowledge-sharing-based method for multi-task security problems.
    In particular, } we modify the state-of-the-art framework in \cite{xu2021deep} and designate the modified method as MTL. \revision{MTL fully utilizes data from both the auxiliary and the target domains, thus learning similar knowledge to \ours.} Specifically, it mixes data from multiple auxiliary domains and the target domain to train a model on the target domain. To prevent the effect of task imbalance (e.g., the target domain has fewer requests), we assign weights to the loss of requests from each domain in accordance with the number of requests from with limited data. 
    \item \textbf{SCNN. } SCNN leverages a specially designed CNN to classify HTTP requests \cite{zhang2017deep}. For better performance, we use the preprocessing technique of ZeroWall to produce token sequence. Since supervised methods require labeled data, we collect attack requests filtered by WAFs from a day for each domain and use them as malicious samples, i.e., their training data contains 1-day known attack requests and benign modify{requests} collected at different lengths of time. 
    \item \textbf{SRNN. } SRNN utilizes the attention-based LSTM to classify sequence data \cite{zhou2016attention}. The settings for preparing training data and preprocessing are the same as SCNN.
\end{itemize}

For unsupervised methods \ours, ZeroWall, and MTL, to make a fair comparison, we choose LSTM as the backbone of encoder-decoder networks \revision{following \cite{16tang2020zerowall}.} LSTM has been demonstrated to be more effective at modeling sequence data than GRU or classic RNNs \cite{cho2014learning, du2017deeplog}. Moreover, \revision{Transformers may learn an identity function that reconstructs HTTP requests including attacks, and thus are ill-suited for the detection methods that identify attacks based on reconstruction errors \cite{vaswani2017attention,ng2011sparse}. Our empirical study shows that, if we replace LSTM with Transformer, the F1-score of \ours significantly drops from 0.913 to 0.487 in domain OV with 5-minute training data. }

\begin{table}[t]
    \centering
    \caption{Statistics on 4 domains. }
    \vspace{-0.1cm}
    \label{tab:datasets}
    \begin{tabular}{l|r|r}
    \toprule[0.8pt]
        \textbf{Domain} & \textbf{$\#$ of Requests} & \textbf{$\#$~of Attacks} \\
    \midrule[0.8pt]
        Online Video (OV) & 57.71M & 392 \\
        Social Network Service (SNS) & 84.72M & 818\\
        Sports Forum (SF) & 13.85M & 2,623\\
        Identity Management (IdM) & 136.78M & 59\\
    \bottomrule[0.8pt]
    \end{tabular}
\end{table}

\begin{table*}[t]
\centering
\caption{Overall performance comparisons when the collection time of training data is 5 minutes and 1 day respectively. }
\vspace{-0.1cm}
\begin{tabular}{m{0.9cm}<{\centering}|r|ccc|ccc|ccc|ccc}
\toprule[0.8pt]
\multirow{2}{*}{\textbf{Time}}   & \multirow{2}{*}{\textbf{Method}} & \multicolumn{3}{c|}{\textbf{OV}} & \multicolumn{3}{c|}{\textbf{SNS}} & \multicolumn{3}{c|}{\textbf{SF}} & \multicolumn{3}{c}{\textbf{IdM}} \\
                        &        & \textbf{Pre} & \textbf{Rec} & \textbf{F1} &
                        \textbf{Pre} & \textbf{Rec} & \textbf{F1} & 
                        \textbf{Pre} & \textbf{Rec} & \textbf{F1} & 
                        \textbf{Pre} & \textbf{Rec} & \textbf{F1} 
                        \\
                        \midrule[0.8pt]
\multirow{5}{*}{5 min}  & ZeroWall                & 0.580    & 0.918    & 0.711    & 0.168    & 0.536    & 0.256    & 0.422    & 0.964    & 0.587    & 0.647    & 0.932    & 0.764   \\
                        & MTL                     & 0.777    & 0.623    & 0.692    & 0.268    & 0.833    & 0.406    & 0.498    & 0.955    & 0.655    & 0.845    & 0.831    & 0.838   \\
                        & SCNN                     & 0.080    & 0.940    & 0.147    & 0.174    & 0.845    & 0.288    & 0.084    & 0.699    & 0.150    & 0.377    & 0.966    & 0.543   \\
                        & SRNN                     & 0.040    & 0.338    & 0.072    & 0.070    & 0.972    & 0.131    & 0.073    & 0.717    & 0.133    & 0.177    & 0.983    & 0.300   \\
                        & Ours                   & 0.901    & 0.926    & 0.913    & 0.667    & 0.911    & 0.770    & 0.685    & 0.972    & 0.802    & 0.868    & 1.000    & 0.929    \\ \midrule
\multirow{5}{*}{1 day}  & ZeroWall                & 0.924    & 0.866    & 0.894    & 0.810    & 0.599    & 0.719    & 0.802    & 0.830    & 0.816    & 0.879    & 0.864    & 0.872   \\
                        & MTL                     & 0.935    & 0.861    & 0.896    & 0.902    & 0.513    & 0.654    & 0.486    & 0.797    & 0.603    & 0.962    & 0.847    & 0.901   \\
                        & SCNN                     & 0.954    & 0.395    & 0.559    & 0.678    & 0.530    & 0.595    & 0.936    & 0.393    & 0.553    & 0.851    & 0.678    & 0.755   \\
                        & SRNN                     & 0.960    & 0.662    & 0.784    & 0.324    & 0.185    & 0.236    & 0.529    & 0.330    & 0.407    & 0.897    & 0.593    & 0.714   \\
                        & Ours                   & 0.921    & 0.884    & 0.902    & 0.895    & 0.819    & 0.855    & 0.922    & 0.893    & 0.907    & 1.000    & 0.915    & 0.956   \\ \bottomrule[0.8pt]
\end{tabular}

\label{tab:main}
\end{table*}

\revision{
\parab{Selection of the Base Domain.} We select a domain among the auxiliary domains as the base domain if the domain has the most overlapping tokens with other domains. It allows us to maximize the overlap for token alignment and thus best utilizes the correlation between different domains for detection. We also conduct an empirical study in Appendix \ref{sec:base_domain} and show that the selection of the base domain has limited impacts on the model performance.
}

\parab{Implementation.} All the evaluations are conducted on a GPU server whose hardware environment is configured as 10-core vCPU, NVIDIA Tesla T10 GPU, and 40GB memory. Deep learning models are implemented in PyTorch 1.7.1 with CUDA 11.0 toolkit.  We use Gensim to implement Word2vec.

\parab{Configurations.} When evaluating the performance of one domain, we use the other three domains as auxiliary domains to perform MTL or cross-domain training of \ours. Note that, since the universal initial models only need to be trained once, they do not suffer from limited data, thus we train them using full datasets from auxiliary domains.
For our method, we use the skip-gram to train the Word2vec embeddings. For the LSTM-based seq2seq model adopted in \ours, ZeroWall, and MTL, 
both the hidden size and embedding size are set to be 512. We use the Adam \cite{kingma2014adam} with the learning rate of 0.001 as the optimizer and the learning rate decay is applied. To make an efficient computation on GPU, we use token-level dynamic batching to train the model. The token batch size is set to 4096.

\parab{Ethical Considerations.} The data that we analyze only contains URLs and bodies of requests that are helpful for detection. Any fields containing sensitive user information have been removed.
We also do not use the collected data to identify any individual. 
All datasets are stored on the company’s servers and are accessed through an internship program. We conduct experiments in an isolated environment that has no impact on the production environment.

\begin{figure}[t] 
\centering 
\begin{subfigure}[b]{0.7\linewidth}
\centering
\includegraphics[width=0.9\textwidth]{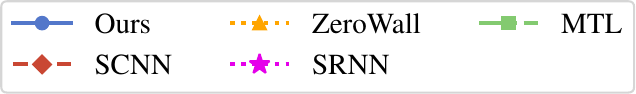}
\end{subfigure}
\hspace{0.2cm}
\begin{subfigure}[b]{0.47\linewidth}
         \centering
         \includegraphics[width=\textwidth, trim=0 5 0 2, clip]{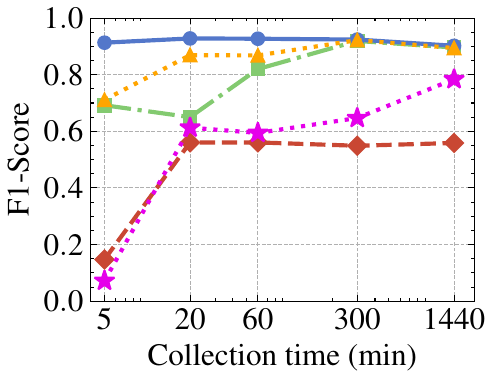}
         \caption{OV}
     \end{subfigure}
     \hspace{0.2cm}
     \begin{subfigure}[b]{0.47\linewidth}
         \centering
         \includegraphics[width=\textwidth, trim=0 5 0 2, clip]{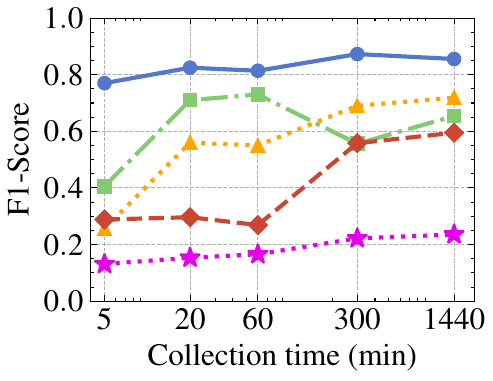}
         \caption{SNS}
     \end{subfigure}
     
     \begin{subfigure}[b]{0.47\linewidth}
         \centering
         \includegraphics[width=\textwidth, trim=0 5 0 2, clip]{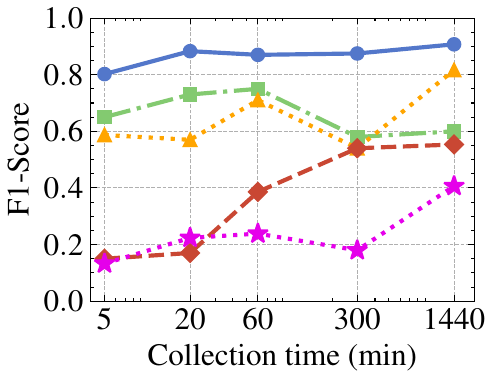}
         \caption{SF}
     \end{subfigure}
     \hspace{0.2cm}
     \begin{subfigure}[b]{0.47\linewidth}
         \centering
         \includegraphics[width=\textwidth, trim=0 5 0 2, clip]{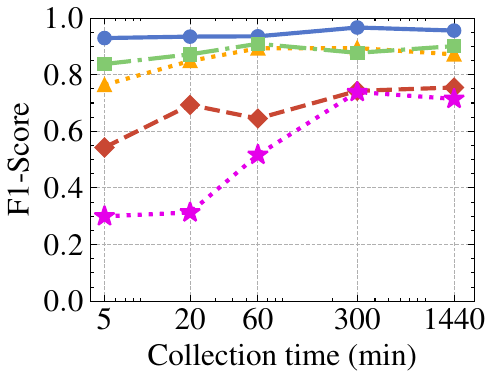}
         \caption{IdM}
     \end{subfigure}

     \vspace{-0.3cm}
\caption{Detection performance with respect to the amount of training data.}
\label{fig:changing} 
\end{figure}

\subsection{Detection performance}

First, we run experiments to evaluate whether our method outperforms baselines with limited training data. We use the 5-minute data for model training and test the trained model on the testing data.
Table \ref{tab:main} shows the detection performance of our method and baselines. It can be seen that our method significantly outperforms baselines when the collection time is 5 minutes. Compared with ZeroWall, our method improves the F1-score by 21.5\% at least. The results confirm the advantages of leveraging knowledge sharing when the training data is limited. We can further observe that such an improvement is mainly derived from precision (e.g., improved by 55.3\% on OV), which \revision{indicates that our method effectively mitigates the false positives caused by the limited training data.} Moreover, MTL can slightly improve model performance when the collection time is 5 minutes. For example, the F1-scores of SNS using MTL are improved from 0.256 to 0.406 compared with ZeroWall. However, this improvement is still far from ours. Overall, our method is more effective with limited Web training data.

Second, 
our method is also better than baselines when the collection time is 1 day. For example, the F1-scores of SNS and IdM are improved by 18.9\% and 9.6\% respectively compared with ZeroWall. This indicates that even with abundant training data, 
our method still helps to improve the generalizability of the trained model by combining the knowledge from other domains. 
Note that the results of MTL are even lower than that of ZeroWall when the collection time is 1 day. We analyze the main reason is that one or more auxiliary domains may have a greater impact than the target domain, which degrades the detection performance on the target domain. \looseness=-1

Third, supervised methods SCNN and SRNN perform the worst when compared to unsupervised methods. For the 5-minute evaluation, both methods are ineffective due to insufficient data. For the 1-day evaluation, their recall is poor compared to their precision, indicating many attacks are undetected. This is because supervised methods depend on prior knowledge of attack requests and thus fail to effectively detect zero-day attacks, which were previously unknown. 
\revision{Conversely, \ours is capable of detecting zero-day attacks because it detects attacks only according to learned benign patterns of Web requests}.

\revision{Fourth, the detection performance varies across domains. For example, 
\ours performs worse on SNS and SF than on OV and IdM.
This is mainly because the services running in different domains invoke different APIs with various parameters, and thus the complexity of 
benign patterns 
of these services varies. It is difficult for the detection model to learn these complicated patterns, which may result in more false positives and lower precision.}

Figure \ref{fig:changing} shows the F1-scores given different collection times of the training data.
\modifyy{We} observe that our method only needs 5-minute training data to reach the performance that the baseline method uses 1-day training data. 
Moreover, for all methods, model performance increases with collection time in most cases. If the collection time is shorter, the performance gap between the baseline method and our method will be larger. However, it is worth noting that more collection time does not always lead to better performance. For example, the F1-score of 1-day training data is slightly lower than that of using 5-minute training data on OV. This performance loss comes from recall (reduced from 0.926 to 0.884) according to the detailed breakup of performance shown in Table \ref{tab:main}, which is mainly because the training data \modifyy{includes} some noise samples that reduce the ability of the unsupervised model to detect zero-day attacks. 
We also discuss this phenomenon in the subsequent section.

We further perform a case study to investigate how our method helps to improve detection performance. We show a benign request that is misjudged by ZeroWall but correctly recognized by our method in IdM (note that some fields are masked for preserving anonymity and privacy) in Table \ref{tab:case-fp}. ZeroWall produces the false positive because the pattern of the token sequence ``\texttt{xxxx\_code \_num\_ bids \_num\_ yyyyy\_code \_num\_}'' is not included in the training data. However, requests with a similar pattern may exist in other domains (Table \ref{tab:case-similar} shows two examples in OV and SNS, respectively), which allows our method to better learn the pattern of these requests and avoid the false positive.

\begin{table}[t]
\centering
\caption{A benign request that is incorrectly recognized by baseline as an attack but is correctly recognized by \ours.}
\vspace{-0.1cm}
\begin{tabular}{m{0.2\linewidth}|m{0.7\linewidth}}
\toprule
\makecell[c]{\textbf{Request}} & \makecell[l]{.../send?xxxx\_code=8**\&bids=1*****76 \\ \&yyyyyy\_code=18***6} \\ \midrule
\makecell[c]{\textbf{Token} \\ \textbf{Sequence}} & \makecell[l]{
... send \textbf{xxxx\_code \_num\_ bids \_num\_}  \\ \textbf{yyyyyy\_code \_num\_}
}\\ \bottomrule
\end{tabular}
\label{tab:case-fp}
\end{table}

\begin{table}[t]
\centering
\caption{Token sequence of two example requests that have a similar structure to that shown in Table \ref{tab:case-fp}. Example 1 and 2 are from OV and SNS respectively.}
\vspace{-0.1cm}
\begin{tabular}{m{0.2\linewidth}|m{0.72\linewidth}}
\toprule
\textbf{Example 1} & \makecell[l]{watch\_record\_new callback \_pbas\_ \textbf{pagesize} \\ \textbf{\_num\_ g\_tk \_num\_ g\_vstk \_num\_}} \\ \midrule
\textbf{Example 2} & \makecell[l]{cgi\_qzshare \textbf{uin \_num\_ spaceuin  \_num\_} \\ \textbf{g\_iframe  \_num\_}}\\ \bottomrule
\end{tabular}
\label{tab:case-similar}
\end{table}

\begin{figure}[t]
    \centering
    \begin{subfigure}[b]{0.75\linewidth}
        \includegraphics[width=\textwidth]{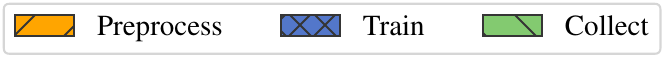}
    \end{subfigure} \\
    \begin{subfigure}[b]{0.9\linewidth}
        \includegraphics[width=\textwidth, trim=0 5 0 0, clip]{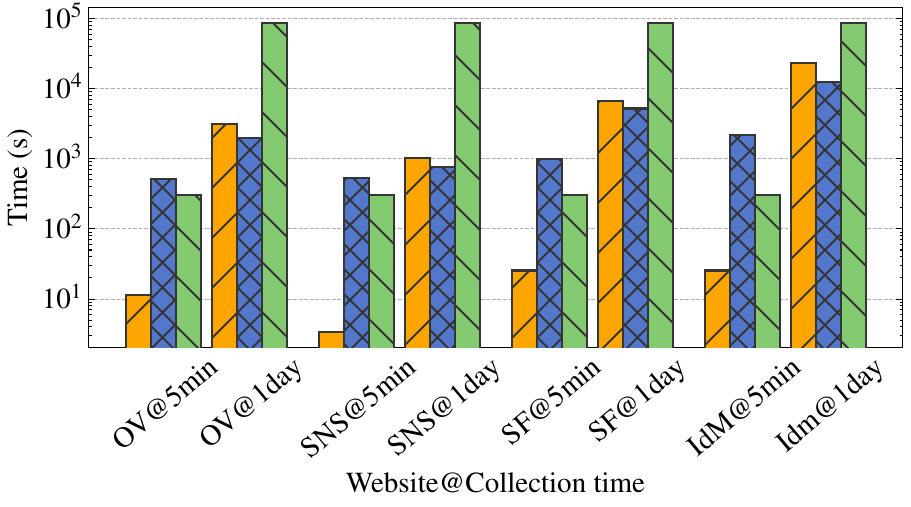}
    \end{subfigure}

    \vspace{-0.1cm}
    \caption{Development times of \ours with 5-minute training data and ZeroWall with 1-day training data.}
    \label{fig:time}
\end{figure}

\subsection{Development Overhead}
We now assess the development overhead of our proposed method. Considering that our proposed method with 5-minute training data achieves comparable detection performance to ZeroWall using 1-day training data, we compare the development times of these two settings. Figure \ref{fig:time} displays the comparison results in terms of collection time, preprocessing time, and training time, where training time refers to the time to train a convergent model. 
It can be seen that \ours is hundreds of times faster than ZeroWall for both collection and preprocessing. Meanwhile, the training time of \ours is only half that of ZeroWall. These results suggest that \ours is more efficient in developing new detection models than ZeroWall, which requires a large amount of training data.

\revision{We also examine the training time for universal initial models with the three auxiliary domains. Training universal initial models of OV, SNS, SP, and IdM takes 437, 487, 459, and 333 minutes, respectively. Note that universal initial models can be prepared in advance, thus not affecting the efficiency of developing new detection models.}

\begin{table*}[t]
\centering
\caption{Comparison with and without our adaptive preprocessing. A.P. is the abbreviation of adaptive preprocessing.}
\vspace{-0.1cm}
\begin{tabular}{m{0.9cm}<{\centering}|r|ccc|ccc|ccc|ccc@{}}
\toprule
\multirow{2}{*}{\textbf{Time}} & \multirow{2}{*}{\textbf{Method}}    & \multicolumn{3}{c|}{\textbf{OV}} & \multicolumn{3}{c|}{\textbf{SNS}} & \multicolumn{3}{c|}{\textbf{SF}} & \multicolumn{3}{c}{\textbf{IdM}} \\
                      &                            & \textbf{Pre}      & \textbf{Rec}      & \textbf{F1}       & \textbf{Pre}      & \textbf{Rec}      & \textbf{F1}       & \textbf{Pre}      & \textbf{Rec}      & \textbf{F1}       & \textbf{Pre}      & \textbf{Rec}      & \textbf{F1}      \\ \midrule
\multirow{2}{*}{5 min}                 & Ours w/o A.P.  & 0.908    & 0.903    & 0.905    & 0.458    & 0.745    & 0.567    & 0.525    & 0.992    & 0.686    & 0.825    & 0.994    & 0.902   \\
                      & Ours                       & 0.901    & 0.926    & 0.913    & 0.667    & 0.911    & 0.770    & 0.685    & 0.972    & 0.802    & 0.868    & 1.000    & 0.929   \\ \midrule
\multirow{2}{*}{1 day}                 & Ours w/o A.P. & 0.919    & 0.876    & 0.897    & 0.745    & 0.714    & 0.729    & 0.763    & 0.893    & 0.823    & 1.000    & 0.847    & 0.917   \\
                      & Ours                       & 0.921    & 0.884    & 0.902    & 0.895    & 0.819    & 0.855    & 0.922    & 0.893    & 0.907    & 1.000    & 0.915    & 0.956   \\ \bottomrule
\end{tabular}

\label{tab:as_token}
\end{table*}

\subsection{Ablation Study}

\begin{table*}[]
\centering
\caption{Comparison of different variants of our method with knowledge sharing. K.S. and T.L. are the abbreviations for knowledge sharing and transfer learning, respectively.}
\vspace{-0.1cm}
\begin{tabular}{m{0.9cm}<{\centering}|r|rcc|ccc|ccc|ccc}
\toprule[0.8pt]
\multirow{2}{*}{\textbf{Time}} & \multirow{2}{*}{\textbf{Method}}    & \multicolumn{3}{c|}{\textbf{OV} }& \multicolumn{3}{c|}{\textbf{SNS}} & \multicolumn{3}{c|}{\textbf{SF}} & \multicolumn{3}{c}{\textbf{IdM}} \\
                      &                            & \textbf{Pre}      & \textbf{Rec}      & \textbf{F1}       & \textbf{Pre}      & \textbf{Rec}      & \textbf{F1}       & \textbf{Pre}      & \textbf{Rec}      & \textbf{F1}       & \textbf{Pre}      & \textbf{Rec}      & \textbf{F1}      \\ \midrule[0.8pt]
\multirow{3}{*}{5 min} & Ours w/o K.S.       & 0.633    & 0.943    & 0.758    & 0.319    & 0.774    & 0.452    & 0.470    & 0.989    & 0.637    & 0.883    & 0.898    & 0.891   \\
                       & Ours w/ T.L.       & 0.431    & 0.926    & 0.588    & 0.389    & 0.806    & 0.525    & 0.456    & 1.000    & 0.626    & 0.850    & 0.864    & 0.857   \\
                       & Ours                   & 0.901    & 0.926    & 0.913    & 0.667    & 0.911    & 0.770     & 0.685    & 0.972    & 0.802    & 0.868    & 1.000    & 0.929   \\ \midrule
\multirow{3}{*}{1 day} & Ours w/o K.S.      & 0.914    & 0.882    & 0.898    & 0.893    & 0.689    & 0.778     & 0.947    & 0.824    & 0.881    & 0.943    & 0.847    & 0.893   \\
                       & Ours w/ T.L.       & 0.899    & 0.876    & 0.887    & 0.844    & 0.620    & 0.715    & 0.975    & 0.696    & 0.812    & 0.981    & 0.898    & 0.938   \\
                       & Ours                    & 0.921    & 0.884    & 0.902    & 0.895    & 0.819    & 0.855    & 0.922    & 0.893    & 0.907    & 1.000    & 0.915    & 0.956   \\ \bottomrule[0.8pt]
\end{tabular}
\label{tab:as_meta}
\end{table*}

To further validate the design of RETSINA, we evaluate how different modules contribute to improving detection performance.
First, we assess how adaptive preprocessing supports semantics analysis. We replace the adaptive preprocessing with the naive tokenization technique commonly used in previous work \cite{16tang2020zerowall}, denoting \textit{ours without adaptive preprocessing}. Table \ref{tab:as_token} shows the comparison of detection performance. 
It can be seen that adaptive preprocessing improves the F1-scores under all settings. The improvements are significant on SNS, SF, and IdM when the collection times are 5 minutes and 1 day. For example, without adaptive preprocessing, the F1-score on SNS drops by 26.3\%, and its corresponding precision and recall drop by 31.3\% and 18.2\%, respectively. Note that, the improvements brought by adaptive preprocessing in OV are minor since the requests in OV are relatively simple and can be mostly handled by naive tokenization techniques. 

Second, we remove the multi-domain representation module and the cross-domain module. 
Note that, we treat the multi-domain representation module and the cross-domain training module as being inseparable in knowledge sharing based on meta-learning because 1) multi-domain representation performs universal embedding mapping without changing the relationship between embeddings of tokens in the same domain, thus training a model using universal embedding from only one domain brings no benefit, and 2) cross-domain training requires inputs to be in the same feature space, which cannot be guaranteed without universal embedding.
Therefore, we consider two variants for these two modules: 1) \textit{ours without knowledge sharing}. We use the token sequences generated by the adaptive preprocessing module to train a seq2seq detection model and 2) \textit{ours with transfer learning}. We leverage the common transfer learning paradigm as an alternative choice for knowledge sharing. A model is pre-trained using data from multiple auxiliary domains and is then fine-tuned to the target domain, where the embeddings of unknown tokens are randomly initialized. Table \ref{tab:as_meta} shows the results. Under all settings, our method achieves the best performance with the multi-domain representation and the cross-domain training.  Moreover, these two modules are especially helpful to boost detection performance when the data is limited. When these two modules are removed, F1-scores drop at most 41.3\% and 9.0\% with 5-minute and 1-day training data, respectively.  Besides, the performance of transfer learning is even worse than without knowledge sharing in most cases, indicating that transfer learning is not effective for sharing knowledge among heterogeneous Web domains. \looseness=-1

\begin{figure}[t] 
\centering 
\begin{subfigure}[b]{0.45\linewidth}
         \centering
         \includegraphics[width=\textwidth,trim=0 3 0 3,clip]{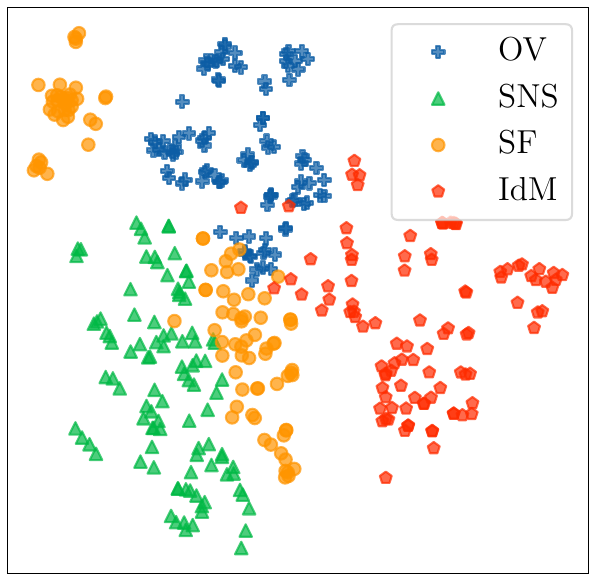}
         \caption{Word2vec}
     \end{subfigure}
     \hspace{0.2cm}
     \begin{subfigure}[b]{0.45\linewidth}
         \centering
         \includegraphics[width=\textwidth,trim=0 3 0 3,clip]{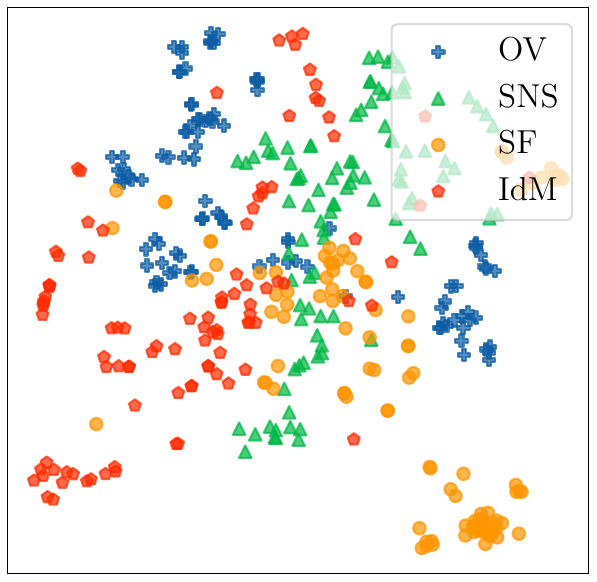}
         \caption{Ours}
     \end{subfigure}

     \vspace{-0.2cm}
\caption{T-SNE Visualization of token embeddings using Word2vec and our method.} 
\label{fig:vis} 
\end{figure}

We further visualize the distribution of token embeddings in different domains to illustrate how our multi-domain representation helps to achieve knowledge sharing from heterogeneous Web data. We randomly select 100 tokens in each domain and project their embeddings into 2-dimension using t-SNE \cite{van2008visualizing}. 
\revision{The embedding of t-SNE is initialized by principal component analysis, which is more globally stable than random initialization \cite{TSNE}.} 
Figure \ref{fig:vis} shows the results using our multi-domain representation and the Word2vec technique. Using Word2vec, embeddings from each domain form a cluster, which indicates the tokens of each domain are in different feature spaces. On the contrary, with multi-domain representation, tokens from the same domain are more dispersed and tokens from different domains are mixed. This suggests that multi-domain representation makes tokens from different domains share the same feature space, and thus the correlation between tokens is captured. 

\begin{figure}[] 
\centering 
\begin{subfigure}[b]{0.7\linewidth}
    \centering
    \includegraphics[width=\textwidth]{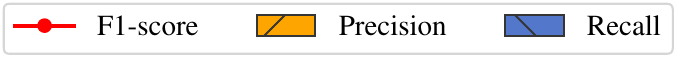}
\end{subfigure}\\

\begin{subfigure}[b]{0.45\linewidth}
         \centering
         \includegraphics[width=\textwidth]{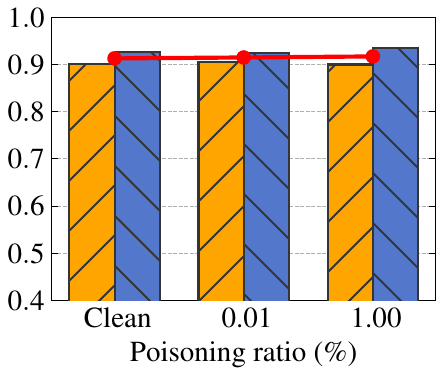}
         \caption{Ours-5 mins}
     \end{subfigure}
     \hspace{0.2cm}
     \begin{subfigure}[b]{0.45\linewidth}
         \centering
         \includegraphics[width=\textwidth]{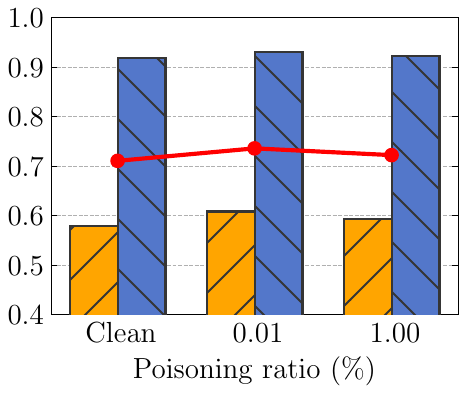}
         \caption{ZeroWall-5 mins}
     \end{subfigure}
     
     \begin{subfigure}[b]{0.45\linewidth}
         \centering
         \includegraphics[width=\textwidth]{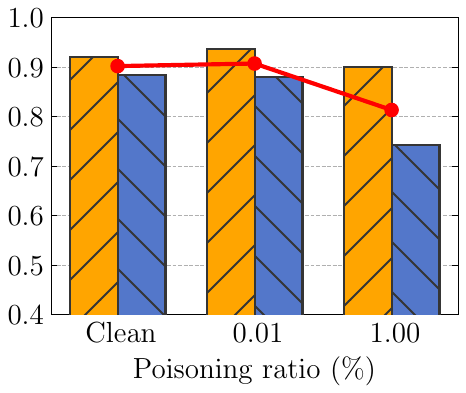}
         \caption{Ours-1 day}
     \end{subfigure}
     \hspace{0.2cm}
     \begin{subfigure}[b]{0.45\linewidth}
         \centering
         \includegraphics[width=\textwidth]{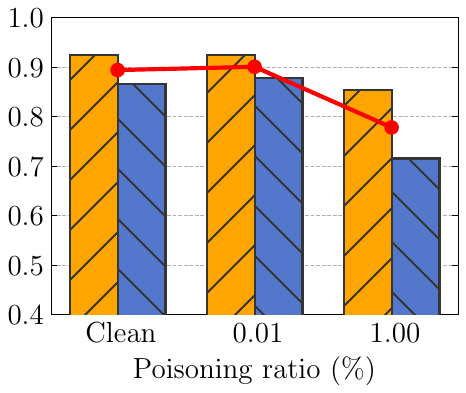}
         \caption{ZeroWall-1 day}
     \end{subfigure}

    \vspace{-0.2cm}
\caption{Comparison of detection performance under data poisoning.}
\label{fig:poison} 
\end{figure}

\begin{figure}[t] 
    \centering 
\begin{subfigure}[b]{0.47\linewidth}
         \centering
         \includegraphics[width=\textwidth]{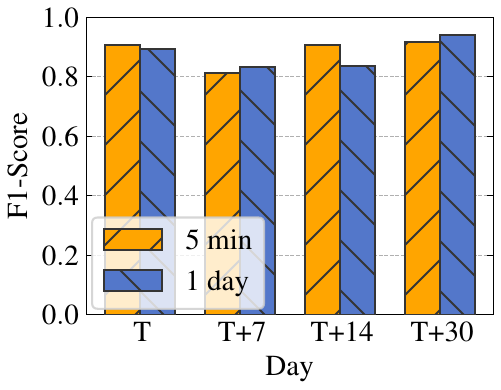}
         \caption{OV}
     \end{subfigure}
     \hspace{0.2cm}
     \begin{subfigure}[b]{0.47\linewidth}
         \centering
         \includegraphics[width=\textwidth]{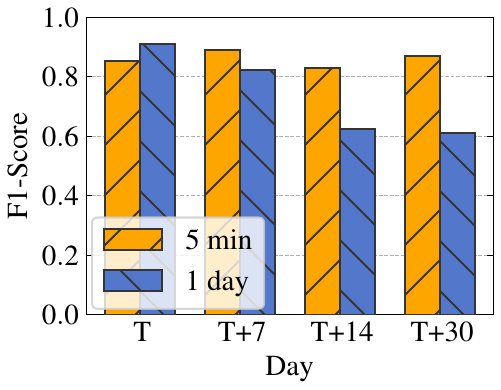}
         \caption{SNS}
     \end{subfigure}

\vspace{-0.2cm}
\caption{Performance of our method in online deployment when the training data is collected taking 5 minutes and 1 day respectively.} 
\label{fig:concept_drift_main} 
\end{figure}

\subsection{Robustness under Data Poisoning} \label{sec:poison}
In this section, we quantify how effective our method is against data poisoning. We realize data poisoning by generating and adding poison samples to the original dataset for model training. Specifically, the poison samples are automatically generated by a commercial Web security scanner\footnote{The scanner presets over 3000 human-written attack instances which cover more than 40 types of common Web vulnerabilities including remote command execution, SQL injection, etc. To generate poison samples, it first collects a set of normal requests and substitutes query values in the HTTP request based on these instances.}. The poison samples are added to the original dataset at different ratios, denoted as poison ratios $\eta$. We vary $\eta$ over [0\%, 0.1\%, 1\%]. We test both our method and ZeroWall using different collection time and present the comparison of detection performance on OV in Figure \ref{fig:poison}.

Overall, our method outperforms ZeroWall when dealing with data poisoning and is more stable. When the collection time is 5 minutes, both methods are stable when the poison ratio increases, but our method provides better detection performance.
When the collection time is 1 day, the detection performance of both methods degrades if we increase the poison ratio to 1\%. We can see that recall drops significantly, compared to precision. This is mainly because the model incorrectly learns the patterns of poisoning samples and identifies malicious samples with similar patterns as benign ones.

Comparing the results of 5-minute and 1-day evaluations, we surprisingly find that, under the same poisoning ratio, models trained with limited data are more robust against data poisoning. This is a result of the fact that, given the same poison ratio, the number of poison samples in 5-minute data is smaller than in 1-day data. To be more specific, when the number of poison samples is small, these samples behave like separate noises and are easily ignored by deep learning models \cite{rolnick2017deep,xu2021differential}. As the number of poison samples increases, it is more probable that these samples will cluster in the feature space and produce a larger norm of gradient that will be captured by the gradient descent algorithm \cite{arpit2017closer}.

\section{Real-World Deployment}
\label{sec:real-world}

To investigate how effective \ours~ is in a real production environment, we deployed it on the domains of OV and SNS in the company. 
The real-world deployment demonstrates the superiority of \ours~ in terms of effectiveness and robustness. Below, we first describe our deployment experience. Then, we present the results of the detection performance and our discoveries.

\subsection{Real-World Deployment Experience}

Since WAFs are effective in detecting known attacks, RETSINA works complementarily with WAFs rather than completely replacing them, to solely concentrate on zero-day attacks. We use the producer-consumer approach to detect online activity. In particular, the WAF acts as the producer, with the detection station, which deploys all of the detection models from each domain, acting as the consumer. Every request permitted by WAF will be mirrored to the detection station for detection. The detection result will be delivered to security operators for additional analysis.

Due to its large volume, real-world traffic is hard for deep learning models to process in a timely manner. 
We thus leverage the hash technique to avoid processing the same token sequence repeatedly. Since requests with the same structure but different valid values are preprocessed into the same token sequence, a hash table is built to store the hash values of token sequences and their detection results for all token sequences that have been processed beforehand. 
For example, the detection model only needs to process one of  ``\texttt{/path/?cid=abcde}'', ``\texttt{/path/?cid=bcdef}'', ``\texttt{/path/?cid=cde\\fg}’’ to return detection results for all three requests since they have the same token sequence ``\texttt{path cid \_cid\_}’.
After the hash tables in the preprocessing significantly reduce the number of token sequences needed to be predicted by the model, the preprocessing itself is the bottleneck for the system throughput. Therefore, we suggest assigning the majority of the CPU resources to the preprocessing concurrently. 
Note that, such a system has good scalability. It can be easily extended by adding multiple backend servers responsible for preprocessing or model prediction in a distributed way.\looseness=-1

\revision{In practice, the universal initial models should be trained before target domain training. Therefore, we use the universal initial models developed in Section \ref{sec:evaluation} that was trained more than 6 months ahead of our real-world deployment to train the target models. The experiment results in Section \ref{sec:real_world_performance} show that the universal initial models are effective for detection, which confirms that the intervals between the data collection for the auxiliary domains and the target domain do not impact the model performance.}

\subsection{Real-World Detection Performance} \label{sec:real_world_performance}
\parab{Detection Results.} In the deployment, we collect \revision{5-minute data on the 0th, 7th, 14th, and 30th days respectively to train the target models.} For OV and SNS, we detect 126 and 218 zero-day \revision{attack requests} on average per day, respectively.
\revision{We also compare the models that are trained using 5-minute or 1-day data, respectively, in Figure \ref{fig:concept_drift_main}.}  
It can be seen that our method can achieve high detection performance using only 5-minute training data, and the F1-scores of 5-minute and 1-day training data are comparable in most cases. The results confirm the effectiveness of our method under limited training data in a real-world environment.

Interestingly, we observe that the results of 1-day experiments can be even worse than those of 5-minute experiments on \revision{the 14th and 30th days} of SNS. We carefully examine the corresponding results and find that, the performance degrades since the training data is poisoned with many attack requests that the WAF is unable to identify. Conversely, the results of the 5-minute experiment behave normally, which is also consistent with the phenomenon observed in Section \ref{sec:poison}.

We also present two examples of zero-day attacks that are detected by our method but missed by WAF in Table \ref{tab:case-zeroday}. Though detected from OV and SNS, respectively, both attacks belong to the same attack type—command injection—where attackers aim to force the server to execute arbitrary codes. 
Benefiting from unsupervised learning, our method detects these attacks based on the learned patterns of benign requests. Conversely, WAFs usually detect command injection attacks based on keyword matching. It is unable to detect these attacks since it does not cover the keyword ``\texttt{\$\%7B@var\_dump}''.

\begin{table}[t]
\centering
\caption{Examples of zero-day attacks in OV and SNS.}
\vspace{-0.1cm}
\begin{tabular}{m{0.12\linewidth}|m{0.75\linewidth}}
\toprule[0.8pt]
\textbf{Domain} & \textbf{Example} \\
\midrule[0.8pt]
\textbf{OV} & \makecell[l]{/***/***?\_=166******782\&callback=\$\%7B@var \\ 
\_dump(md5(158477632))\%7D\&channdlId=0} \\ \midrule
\textbf{SNS} & \makecell[l]{/***/***?desc=***\&md=1\&origin=***\&pics \\ 
=***\&qc=\$\%7B@var\_dump(md5(27365736\\ 
8))\%7D;\&showcount=0\&site=\&summary=*\\ 
**\&url=https://***.com/***\&where=10}\\ 
\bottomrule[0.8pt]
\end{tabular}
\label{tab:case-zeroday}
\end{table}

\parab{Detection under Concept Drift.} 
Web applications are frequently updated, as was earlier mentioned. After the update, concept drift—the phenomenon that the detection performance of a trained model degrades when presented with previously unseen data due to the data update—occurs. We now explore the effectiveness of our approach against concept drift.

\begin{figure}[] 
\centering 
\begin{subfigure}[b]{0.47\linewidth}
         \centering
         \includegraphics[width=\textwidth]{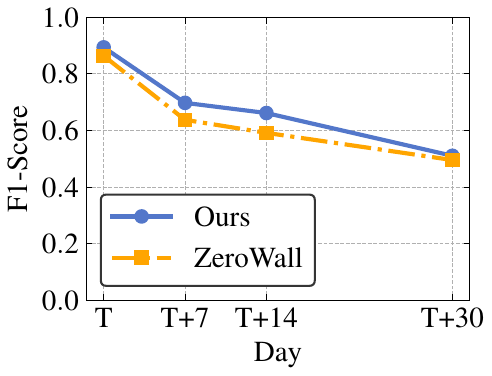}
         \caption{OV}
     \end{subfigure}
     \hspace{0.2cm}
     \begin{subfigure}[b]{0.47\linewidth}
         \centering
         \includegraphics[width=\textwidth]{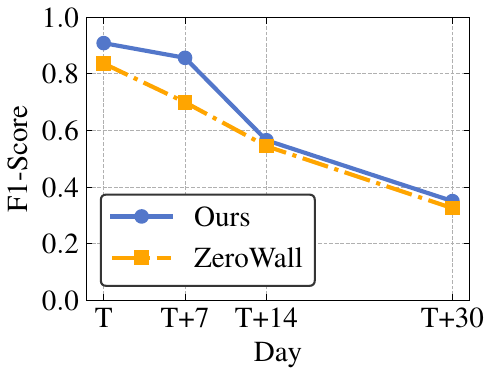}
         \caption{SNS}
     \end{subfigure}

\vspace{-0.2cm}
\caption{Performance comparisons between our method and ZeroWall against concept drift.} 
\label{fig:concep_drift_baseline_compare} 
\end{figure}

Firstly, we examine how RETSINA and the baseline perform against concept drift. 
\revision{We train models using the whole data from \revision{the 0th day} and use them for subsequent testing without retraining.}
Figure \ref{fig:concep_drift_baseline_compare} shows the comparison results. Overall, the detection performance of both our method and ZeroWall degrades. However, our method is more robust to concept drift and always performs better than ZeroWall. For example, for SNS, the F1-score drops by 0.052 in 7 days after training. In comparison, the decrease of ZeroWall is 0.137. This demonstrates that knowledge sharing enables our model to have better generalizability against concept drift. 
\revision{We also examine the corresponding results and find that concepts drift occurs when the Web services are updated, e.g., new APIs are added. Requests that invoke these new APIs are unknown to the old model, resulting in false positives.} 

\begin{figure}[] 
\centering 
\begin{subfigure}[b]{0.47\linewidth}
         \centering
         \includegraphics[width=\textwidth]{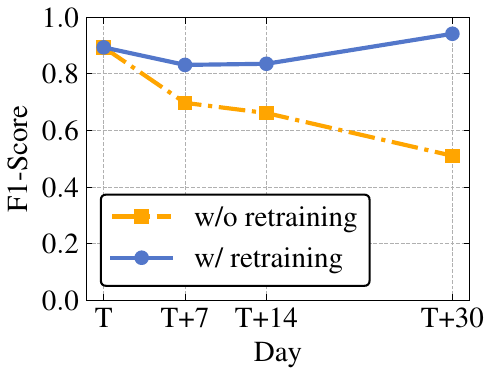}
         \caption{OV}
     \end{subfigure}
     \hspace{0.2cm}
     \begin{subfigure}[b]{0.47\linewidth}
         \centering
         \includegraphics[width=\textwidth]{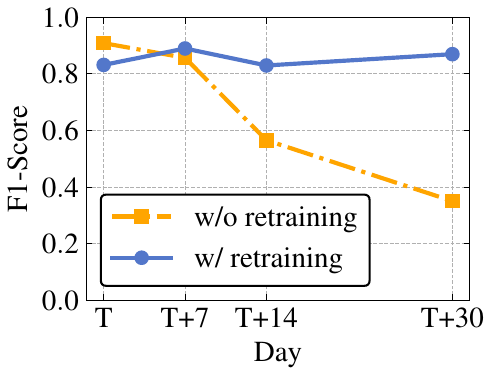}
         \caption{SNS}
     \end{subfigure}
\vspace{-0.2cm}
\caption{The impact of retraining on detection performance.} 
\label{fig:concep_drift_retrain} 
\end{figure}

Secondly, we investigate whether periodic retraining with 5-minute data provides better detection performance than without retraining (i.e., using the model trained on \revision{the 0th day}). Figure \ref{fig:concep_drift_retrain} shows the comparison results. In a word, the results of detection with retraining are stable over time and better than the results without retraining, which validates the necessity of updating the model for preserving the detection performance over time. On the contrary, detection performance degrades severely over time without retraining. For OV, the degradation is significant from \revision{the 7th day}. For SNS, although the performance drop appears to be relatively small on \revision{the 7th day}, the model performance still suffers a huge drop on \revision{the 14th day}. Thus, for these two domains, it is necessary to update the model at least once a week.

\section{Discussion}
\label{sec:discussion}

\parab{Security against Adaptive Attacks.} Given that \ours is a deep learning-based, adversaries may leverage two common techniques for adaptive attacks against deep learning models, i.e., adversarial examples \cite{yuan2019adversarial} and data poisoning \cite{munoz2017towards}, to bypass our detection. We argue that these two techniques are less practical for our framework. 
For adversarial examples, existing works mainly study their impact on supervised classification tasks in computer vision \cite{carlini2017towards} and natural language processing \cite{li2018textbugger}. As discussed in earlier work, their extension to unsupervised tasks only performs with clustering methods \cite{chhabra2020suspicion,yang2020adversarial} or vanilla auto-encoders \cite{goodge2020robustness,gondim2018adversarial}. 
It is still challenging to construct adversarial examples for HTTP requests in practice. First, it is impractical to construct adversarial samples in the input space of our detection model, which is discrete, since HTTP requests are treated as sequential data in \ours. The earlier research on inverse feature-mapping problem shows that attacks are still impractical even when an adversary has white box access to the model and constructs adversarial examples in the feature space through optimization. Second, it is hard to generate adversarial samples while preserving the malicious semantics, and no existing constraint functions measure the similarity of HTTP requests.

Data poisoning requires a certain number of poisoned requests that can bypass the existing WAF, which has a cost. Generally, the number of malicious requests unfiltered by WAFs is relatively small and thus these requests have less impact on the unsupervised model that learns dominated patterns in training data.
Actually, the datasets we evaluate are also directly collected from WAFs and are not completely clean, but as shown in Table \ref{tab:main} our method still achieves high detection performance across all settings. 
This validates that our method is not vulnerable to the small number of attack requests unfiltered by WAFs. 
Moreover, in section \ref{sec:poison}, we show that the model is more robust to poisoning when the number of training data is less given the fixed poison ratio. Recall that learning with limited training data is exactly the strength of our method. We also show that when the collection time of training data is 5 minutes our method will not be affected even if the poisoning ratio is 1\%. All these results demonstrate the robustness of our method against data poisoning. We acknowledge that adversaries may launch more powerful attacks, e.g., \cite{huang2021data,shafahi2018poison}, where only a small number of carefully crafted samples can drastically degrade the model performance.
However, to the best of our knowledge, no related work has been proposed  for the system of Web attack detection. We leave it to future work.

\parab{Limitations.} \ours has the following limitations. First, our method can only judge whether a request is an attack, but cannot provide finer-grained detection results. Thus, it still requires security operators to review the reported attacks. In future work, we plan to incorporate with clustering algorithm and perform fine-grained detection by automatically correlating the attack requests that have similar semantics.
Second, we detect attacks by inspecting the payload of a single HTTP request, which cannot cover the context-based attacks, such as challenge collapsar (CC) attacks \cite{liu2020challenge}. In future work, contextual information can be incorporated into our system to detect these attacks.

\section{Related Work}
\label{sec:related_work}

\parab{Web Attack Detection.} 
\revision{Deploying WAFs \cite{prandl2015study,ModSecurity,WebKnight,Naxsi} is the commonly used way to detect Web attacks in the industry. However, these methods are rule-based (i.e., non-ML based) and cannot detect zero-day attacks that are not matched by rules.} Recently, machine learning based methods have been proposed to improve detection performance. 
These methods work in a supervised or unsupervised manner. Supervised methods \cite{6wang2017detecting,8liang2017anomaly,7liu2022deep, 10das2020network, 11gniewkowski2021http2vec,9liu2019locate,zhang2018anomaly} let the classifier learn to discriminate malicious requests based on previously labeled data. However, due to the huge amount of Web traffic and the rarity of attacks, it is not practical to manually label Web data. In addition, supervised methods assume the knowledge about attacks during training, which can produce unreliable predictions when faced with zero-day attacks \cite{li2022unsupervised,hendrycks2016baseline}.

Unlike supervised methods which learn from both benign requests and malicious requests, unsupervised methods only learn the profile of benign requests, and requests which deviate from the profile are identified as being malicious.
Several methods explore the probability of using the variants of auto-encoder \cite{12vartouni2018anomaly,13park2018anomaly,16tang2020zerowall}. 
Vartouni \textit{et al.} use character-level N-gram and apply stacked auto-encoder \cite{12vartouni2018anomaly}. However, the classification model (i.e., isolation forest) directly uses the output of the encoder which is insufficient for request representation, thus providing limited performance. 
Park \textit{et al.} propose to use a convolutional auto-encoder with character-level binary image transformation \cite{13park2018anomaly}. Tang \textit{et al.} formulate the detection problem as a self-translation problem and apply recurrent seq2seq neural networks \cite{16tang2020zerowall}. 
Our proposed solution also falls into the category of unsupervised methods \revision{and is designed to complement existing WAFs to detect zero-day attacks}. However, in contrast to all methods above, our proposed detection model only requires limited training data and has good generalizability utilizing the knowledge sharing.

\parab{Machine Learning in Multi-task Scenario.} Recently, a series of approaches have been proposed for multi-task learning in different fields \cite{zhang2021survey,misra2016cross,sogaard2016deep,huang2018automatic,xu2021deep}. A common goal of these researches is to learn a unified model to solve a collection of related tasks at the same time. Their key intuition is that learning one task can help improve the performance of other tasks. However, multi-task learning is not a good paradigm to solve our Web attack detection problem because Web applications may be updated frequently. More specifically, we must retrain the unified model when a domain is updated or a new domain comes, which is inefficient compared to updating the model for a single domain name. Moreover, we have compared our method with multi-task learning in our evaluations and demonstrated our advantages.

\parab{Machine Learning with Limited Training Data.} Many works focus on improving the performance of machine learning when using limited training data \cite{tseng2020cross,snell2017prototypical,jiang2018importance,jan2020throwing}. These methods can be grouped into two categories, i.e., data augmentation based and knowledge sharing based. 
Traditional augmentation methods are developed based on domain-specific expertise \cite{simonyan2014very,zhong2020random,kobayashi2018contextual,wei2019eda}. Some new works design algorithms to analyze the feature distribution of available data and generate new data by linear \cite{zhang2017mixup,li2020random,inoue2018data} or non-linear \cite{islam2021crash,jan2020throwing,zheng2019one,perez2017effectiveness,calimeri2017biomedical} interpolations. However, to the best of our knowledge, none of these methods are applicable to generating HTTP request data. 
Knowledge sharing based approaches aim to leverage existing knowledge to boost the model performance on limited training data. Depending on the application scenario, different techniques are developed to achieve knowledge sharing, such as transfer learning \cite{cirecsan2012transfer,wang2015transfer,gu2018universal}, self-supervised learning \cite{devlin2018bert,liang2021fare,liu2021learning} and meta-learning \cite{gu2018meta,zheng2020learning}. Our method falls into this category. The key difference is that we propose specific and novel designs to better capture the correlation of HTTP requests in different domains.

\section{Conclusion}
\label{sec:conclusion}
We propose a novel framework, \ours, to detect zero-day Web attacks across multiple domains with limited training data by utilizing meta-learning. 
We develop a series of new designs to achieve knowledge sharing effectively. 
We evaluate \ours on 4 real-world datasets and demonstrate the effectiveness of \ours for multiple heterogeneous Web domains with limited training data.





\bibliographystyle{ACM-Reference-Format}
\bibliography{ref}

\appendix

\section{Details for Adaptive Preprocessing}
\label{sec:cs_ap}

\begin{figure}[t]
    \centering
    \includegraphics[width=0.95\linewidth]{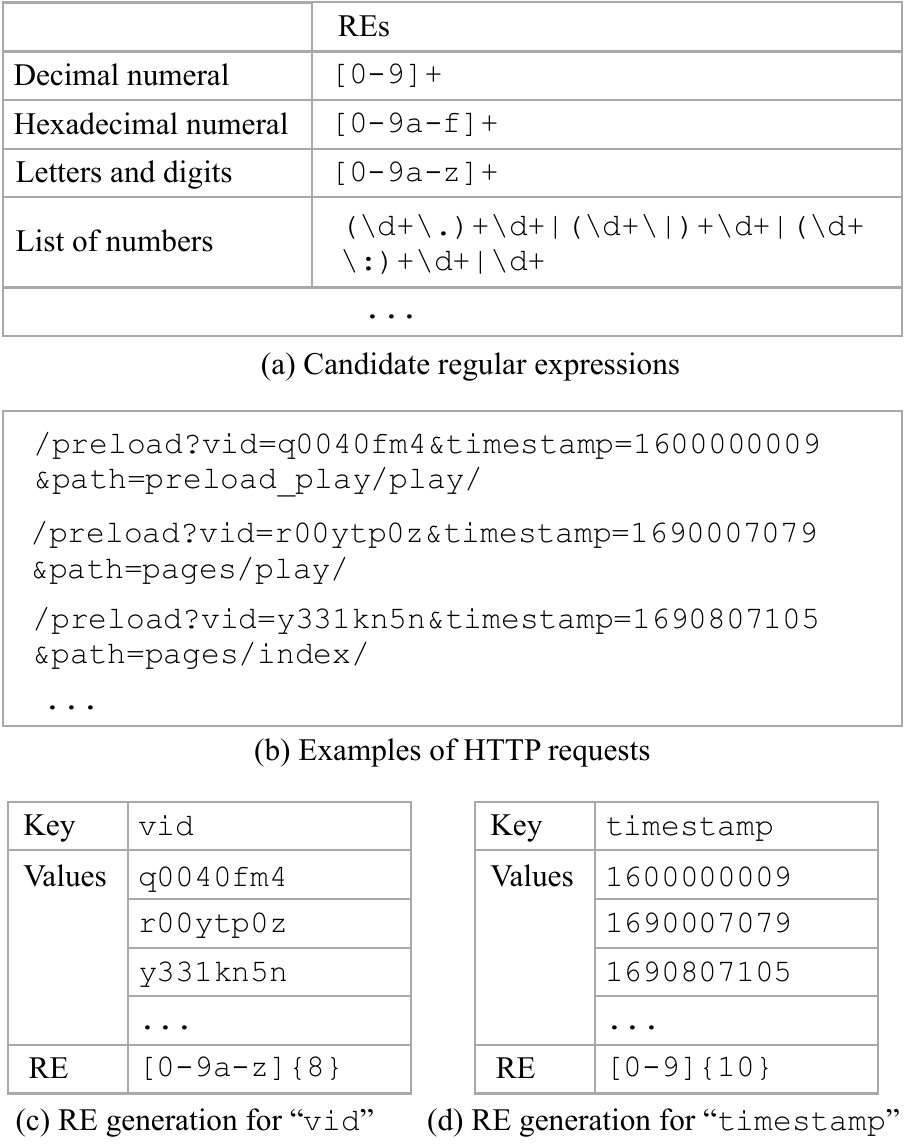}
    \caption{\revision{Examples of regular expression (RE) generation. (a) lists several candidate regular expressions. (b) lists three original HTTP  requests. (c) and (d) show the values and the generated regular expressions of keys ``\texttt{vid}'' and ``\texttt{timestamp}'', respectively.} }
    \label{fig:re-gen}
\end{figure}

\begin{figure}[t] 
    \centering 
    \includegraphics[width=0.95\linewidth]{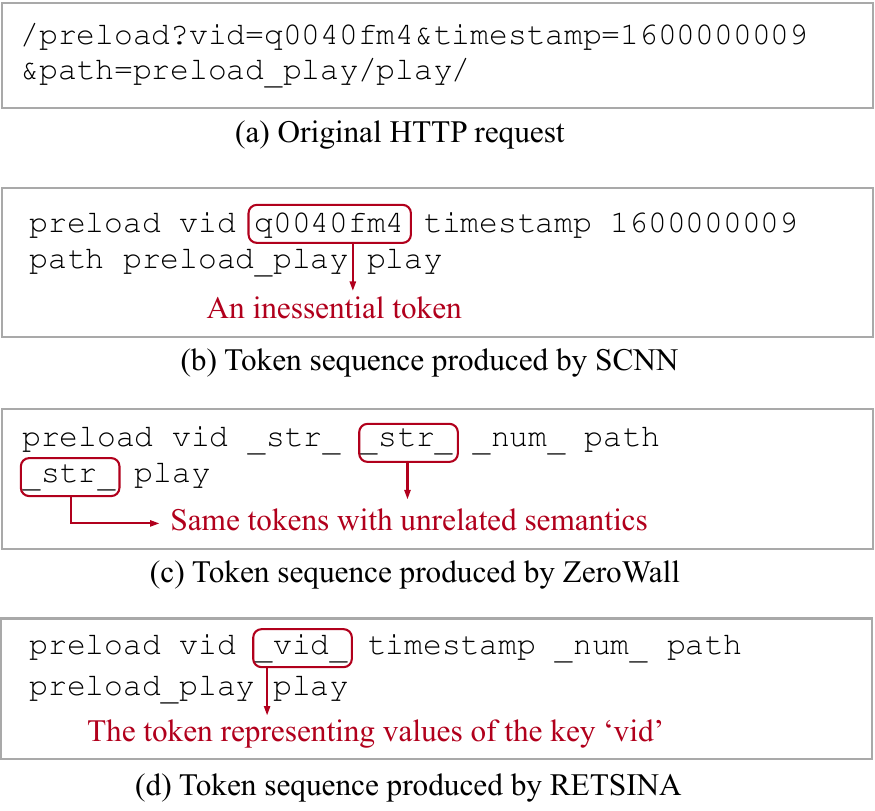}
    \caption{\revision{Comparison of different preprocessing techniques.}} 
\label{fig:cs_ap} 
\end{figure}

\revision{We first illustrate details about generating regular expressions and merging tokens. Then we conduct a case study to demonstrate how our adaptive preprocessing differs from existing works. }

\revision{We generate regular expressions by the following steps. First, given each key and its corresponding values, we enumerate all candidate regular expressions matching these values. Figure \ref{fig:re-gen} (a) shows several candidate regular expressions we devise, which cover the majority of distinct values with inessential information. Note that, letters have been converted to lowercase before regular expression generation. Second, we select the smallest regular expression among the matched regular expressions, which allows us to best extract accurate semantics. Third, we also constrain the repetition of patterns in the regular expression according to the length of values. 
If values are of the same length $x$, we set a common length $x$ (e.g., ``\texttt{[0-9a-z]\{7\}}''). Otherwise, we set a minimum length $x$ such that the lengths of all values are not smaller than $x$ (e.g., ``\texttt{[0-9a-z]\{11,\}}'').} 
\revision{Note that, we relax the constraint of the proportion of values that can be matched by a generated regular expression, instead of requiring the regular expression to match all values, which can eliminate the impact of noises in the requests.} 

\revision{Generally, after obtaining the regular expression for a key, we merge the values of the key that match the regular expression using a placeholder ``\texttt{\_\{key\}\_}'', where ``\texttt{\{key\}}'' represents the name of this key.
Especially when the generated regular expression of a key represents a decimal numeral or a hexadecimal numeral, its values share similar semantics across different domains. Therefore, its values will be merged using the same placeholders, e.g., ``\texttt{\_hexnum\_}''.}

\revision{We now give examples of regular expression generation. Figure \ref{fig:re-gen} (b) shows three examples of HTTP requests in a domain, where ``\texttt{vid}'' represents the identification number of items and ``\texttt{timestamp}'' represents the timestamp of requests. The number of distinct values of ``\texttt{vid}'' or ``\texttt{timestamp}'' could be exceedingly large. Several values and generated regular expressions of key ``\texttt{vid}'' and ``\texttt{timestamp}'' are presented in Figure \ref{fig:re-gen} (c) and (d), respectively. 
For key ``\texttt{vid}'', all its values only match ``\texttt{[0-9a-z]+}'' among all candidate regular expressions and have a length of 8. Thus the regular expression of ``\texttt{vid}'' is ``\texttt{[0-9a-z]\{8\}}''.
For key ``\texttt{timestamp}'', it can be seen that all its values match the first three candidate regular expressions, and all have a length of 10. We pick ``\texttt{[0-9]+}'' since it has the smallest set of elements (i.e., 0-9). Thus we generate ``\texttt{[0-9]\{10\}}''.
}

Our adaptive preprocessing merges tokens in the consideration that the removal of redundant information for HTTP requests helps better extract accurate semantics. Therefore, from the aspect of token merging, we categorize the preprocessing techniques in existing works into two types: i) preprocessing that considers only specific cases of token merging \cite{zhang2017deep,13park2018anomaly,12vartouni2018anomaly} and ii) preprocessing with general strategies for token merging \cite{8liang2017anomaly, 16tang2020zerowall}. Among our baselines, SCNN and ZeroWall are representative methods of these two categories respectively and we compare our method with them. 

\revision{Figures \ref{fig:cs_ap} (b)-(d) present the token sequences produced by different preprocessing methods given the first HTTP request in Figure \ref{fig:re-gen} (b).}  
Since type I preprocessing does not take into account the removal of redundant information from HTTP requests or only considers simplistic cases, they may result in a large number of tokens, such as \texttt{q0040fm4} and other possible ``\texttt{vid}'' values. These inessential tokens would degrade the detection performance due to the data sparsity issue \cite{peng2017addressing}.
Type II preprocessing leverages simple but generic strategies for preprocessing, such as replacing all strings of length over 8 with the same token. This strategy can merge different values of ``\texttt{vid}'', but it may also incorrectly replace other tokens, such as ``\texttt{timestamp}'' and ``\texttt{preload\_play}''. 
While it is possible to limit the replacement to only the ``\texttt{vid}'' field based on expert experience, such an approach is not extensible.
Instead, our method addresses the above issues by automatically finding merging strategies for each key, which helps to identify and remove inessential tokens in a more systematic and scalable way. Our preprocessing recognizes that most of the values of ``\texttt{vid}'' are strings of length 8 and replaces these strings with the same token.

\begin{table}[]
\caption{\revision{Performance comparisons when choosing different base domains. The symbol ``*'' denotes that the base domain is selected using our method.}}
\begin{tabular}{@{}c|c|c|lll@{}}
\toprule
\multicolumn{1}{l|}{\makecell[c]{\textbf{Target} \\ \textbf{Domain}}} & \multicolumn{1}{l|}{\textbf{Time}} & \makecell[c]{\textbf{Base} \\ \textbf{Domain}} & \textbf{Pre}   & \textbf{Rec}   & \textbf{F1}             \\ \midrule
\multirow{6}{*}{OV}   & \multirow{3}{*}{5 min}    & SNS         & 0.896 & 0.929 & 0.912          \\
                      &                           & SP          & 0.890 & 0.921 & 0.905          \\
                      &                           & * IdM        & 0.901 & 0.926 & \textbf{0.913} \\ \cmidrule(l){2-6} 
                      & \multirow{3}{*}{1 day}    & SNS         & 0.932 & 0.885 & \textbf{0.908} \\
                      &                           & SP          & 0.919 & 0.853 & 0.885          \\
                      &                           & * IdM        & 0.921 & 0.884 & 0.902          \\ \midrule
\multirow{6}{*}{IdM}  & \multirow{3}{*}{5 min}    & * OV         & 0.868 & 1.000 & \textbf{0.929} \\
                      &                           & SNS         & 0.895 & 0.864 & 0.879          \\
                      &                           & SP          & 0.889 & 0.949 & 0.918          \\ \cmidrule(l){2-6} 
                      & \multirow{3}{*}{1 day}    & * OV         & 1.000 & 0.915 & \textbf{0.956} \\
                      &                           & SNS         & 0.948 & 0.932 & 0.940          \\
                      &                           & SP          & 0.963 & 0.881 & 0.920          \\ \bottomrule
\end{tabular}
\label{tab:base_domain}
\end{table}

\revision{
\section{Selection of the Base domain }
\label{sec:base_domain}
We evaluate the impact of the selection of base domains on domain OV and IdM. Specifically, we change base domains and compare their performance. The results are shown in Table \ref{tab:base_domain}. It can be seen that the selection of base domains has limited impacts on the model performance, and the fluctuation of F1-scores is within 0.05. Moreover, our method selects the base domain with the best performance in most cases. 
}
\end{document}